\documentclass[a4paper,12pt]{article}
\usepackage{epsfig}
\usepackage[utf8]{inputenc}
\usepackage{amsmath,amssymb}
\usepackage{graphicx}
\usepackage{mathtext}
\usepackage[sort, numbers]{natbib}

\newcommand{\be}{\begin{equation}}
\newcommand{\ee}{\end{equation}}
\newcommand{\bea}{\begin{eqnarray}}
\newcommand{\eea}{\end{eqnarray}}
\begin{document}

\title{Strong shock in the uniformly expanding universe with a spherical void}

\author{G.S. Bisnovatyi-Kogan$^{1, 2, 3}$\thanks{Email: gkogan@iki.rssi.ru},
S.A. Panafidina$^{1,3}$\thanks{Email: sofya.panafidina@phystech.edu}\\
{\small $^{1}$Space Research Institute RAS, Moscow, Russia};\\
{\small $^{2}$National Research Nuclear University MEPhI, Moscow, Russia};\\
{\small $^{3}$Moscow Institute of Physics and Technology MIPT, Moscow reg., Russia}
}

\date{}
\maketitle

\begin{abstract}

Propagation of strong shock wave in the expanding universe is studied using approximate analytic, and exact numerical solution
of self-similar equations. Both solutions have similar properties, which change qualitatively, depending on the adiabatic powers $\gamma$.
In the  interval $1<\gamma<\gamma_{cr} \sim 1.16$  analytic and numeric solutions fill all the space without any voids
and they are rather close to each other. At larger $\gamma>\gamma_{cr}$ a pressure becomes zero at finite radius, and a spherical void appears
around the origin in both solutions. All matter is collected in thin layer behind the shock wave front. The structure of this layer
 qualitatively depends on $\gamma$. At the inner edge of the layer  the pressure is always zero, but the density on this edge
is jumping from zero to infinity at $\gamma \approx 1.4$ in both solutions.

\end{abstract}

{\it Keywords:} cosmology, strong shock wave, self-similar solution

\section{Introduction}

Strong explosions could happen at stages of star and galaxy formation, and at last stages of evolution of very massive primordial stars. Observations of GRB optical afterglows have shown existence of heavy elements in the universe at
red shifts up to $z\sim 10$, like in GRB090423 at $z\approx  8.2$,  GRB120923A at  $z\approx 8.5$, GRB090429B with a photo-$z \approx 9.4$ \cite{lzgrb}. The heavy elements should be formed in the explosions at earlier stages, at larger red shifts.
Strong explosions are accompanied by formation of a strong shock wave, propagating in the expanding universe. For a static media propagation of strong shocks was was studied by many authors, see e.g.  \cite{stanyuk},\cite{taylor}. Exact analytic solution of self-similar equations, describing strong shock propagation was obtained by L.I. Sedov \cite{sedov,sedov1}. Similar analytic solution was obtained in \cite{bk15}, for a strong explosion in the expanding media of a flat Friedman dust universe \cite{znuniv}. Contrary to the static media, which has a real zero energy density in the undisturbed state, the zero energy density in the flat Friedman dust universe, in Newtonian approximation, is the result of a sum of the positive kinetic, and negative gravitational energies. This balance cold be lost behind the shock, therefore the analytic solution obtained using the integral of motion similar to \cite{sedov}, is an approximate one. Here we obtain approximate analytic, and exact numerical solutions for the strong shock propagation for a gas at different adiabatic
powers $\gamma$.
 
 It was obtained that numerical solutions, where matter fills the whole space, exist only at $\gamma<\gamma_{cr}=\gamma_* \approx 1.155$. Similar properties are expressed by the approximate analytic solutions with $\gamma_{cr}=\gamma_{*}\approx 1.178$.

    The problem of a strong shock propagation in the expanding medium was considered earlier in different approximations in \cite{be83}- \cite{vob85}. Review of papers on this topic is given in \cite{omke88}.
Propagation of a detonation wave in the flat expanding universe was studied in \cite{kazh86,be85}. Shock propagation in the outflowing stellar wind was considered in \cite{cd89}.

Detailed analysis of solutions with $\gamma>\gamma_{cr}$ revealed a fundamentally difference of the structure of a thin layer near the shock. The pressure at the inner edge of the layer is zero, but density is changing from zero to infinity when $\gamma$ reaches the value $\gamma=\gamma_{cr1}\approx 1.4$. It is the same within numerical errors in numerical and analytical solutions, while the density inside this layer has a quite different behaviour.

\section{Self-similar equations for a strong shock in a uniform expanding medium }

Equations  describing in the Newtonian approximation, a uniformly expanding $v=H(t)r$, self-gravitating medium, with  a density $\rho(t)$ depending only on time, corresponding to the Friedman model of the universe,
 in spherical coordinates is written as  \cite{znuniv}
 \be
 \frac{\partial v}{\partial t}+v \frac{\partial v}{\partial r}=-\frac{1}{\rho} \frac{\partial p}{\partial r}-\frac{G_g m}{r^2},
 \quad \frac{\partial \rho}{\partial t}+ \frac{\partial \rho v}{\partial r}+\frac{2\rho v}{r}=0,
\label{eq1a}
\ee
$$
\quad
\left(\frac{\partial }{\partial t}+v \frac{\partial }{\partial r}\right)\ln{\frac{p}{\rho^\gamma}}=0,
\quad \frac{\partial m}{\partial r}={4\pi}\rho r^2,
$$
where $G_g$ is the gravitational constant.
We consider a flat dusty model  with a
zero velocity at time infinity, having a density $\rho_1(t)$, and expansion velocity $v_1=H_1(t)r$. The solution of the system (\ref{eq1a}) in these conditions is written as
\bea
\rho_1=\delta/t^2,\quad \delta=\frac{1}{6\pi G_g}, \quad \rho_1=\frac{1}{6\pi G_g t^2}; \qquad H_1=\frac{2}{3t}, \quad v_1=2r/3t;\nonumber \\
  m=\frac{4\pi}{3}\rho r^3=\frac{2r^3}{9 G_g t^2},\quad  \frac{G_g m}{r^2}=\frac{2}{9}\frac{r}{t^2}.\qquad\qquad\qquad.
\label{eq20a}
\eea
The Newtonian solution  is physically relevant in the region where
$v_1\ll c_{\rm light}$, $c\ll c_{\rm light}$.
In the case of a point explosion with the energy $E$, at $t=0$,
the number of parameters  is the same as in the static medium
($\delta, \,\,\, E$), therefore
we may look in this case for a self-similar solution in the problem of a strong shock propagation.
 The non-dimensional combination in this case is written as $r(\delta/Et^4)^{1/5}$. A position of the shock in the self-similar solution
corresponds to the fixed value of the self-similar coordinate. The distance of the shock to the center $R$ is written as
 \be
 R=\beta\left(\frac{E t^4}{\delta}\right)^{1/5},
 \label{eq21a}
 \ee
 where $\beta$ is a parameter depending only on the adiabatic power $\gamma$. The velocity of the shock $u$ in the static laboratory frame is written as
\be
 u=\frac{dR}{dt}=\frac{4R}{5t}=\frac{4\beta E^{1/5}}{5\delta^{1/5} t^{1/5}}.
\label{eq22a}
 \ee
The shock propagation velocity $u$, the velocity of the matter behind the shock $v_2$, in the uniformly expanding
medium (\ref{eq20a}), are decreasing with time
$\sim t^{-1/5}$, the pressure behind the shock $p_2$ is decreasing $\sim t^{-2/5}$,
which is  slower than in the case of the constant density
medium. It occurs due to the fact, that the background density is decreasing with time, and the resistance to the shock propagation is decreasing also.

Conditions on the strong shock discontinuity (Hugoniot relations) has the following view

\be
 v_2=\frac{2}{\gamma+1}u+\frac{\gamma-1}{\gamma+1}v^{sh}_1,\,\, \rho_2=\frac{\gamma+1}{\gamma-1}\rho_1,\,\,
 \label{eq23a}
 \ee
$$ p_2=\frac{2}{\gamma+1}\rho_1 (u-v^{sh}_1)^2,\,\,
 c_2^2=\frac{2\gamma(\gamma-1)}{(\gamma+1)^2}(u-v^{sh}_1)^2,
 $$
where $v_1^{sh}=\frac{2R}{3t}$ is the unperturbed expansion velocity on the shock level. The subscript "2" is related to the values behind the shock.
Introduce non-di\-men\-si\-o\-nal variables behind the shock as
 \be
 v=\frac{4r}{5t} V,\,\,\, \rho=\frac{\delta}{t^2} G,\,\,\, c^2=\frac{16r^2}{25 t^2} Z,\,\,\,m=\frac{4\pi}{3}\rho_1 r^3 M=\frac{4\pi}{3}\frac{r^3}{t^2}\delta M,
 \label{eq24a}
 \ee
depending on the self-similar variable $\xi$, written as
 \be
\xi= \frac{r}{R(t)}= \frac{r}{\beta}\left(\frac{\delta}{Et^4}\right)^{1/5}.
 \label{eq25a}
 \ee
In non-dimensional variables (\ref{eq24a}), the conditions (\ref{eq23a}) on the strong shock at $r=R$, $\xi=1$, are written as
\be
V(1)=\frac{5\gamma+7}{6(\gamma+1)}
, \,\,\, G(1)=\frac{\gamma+1}{\gamma-1},\,\,\,
Z(1)=\frac{\gamma (\gamma-1)}{18(\gamma+1)^2},\,\,\,M(1)=1,
\label{eq26a}
\ee
and the  system (\ref{eq20a}) is written as
\be
Z\left(\frac{d\ln Z}{d\ln\xi}+\frac{d\ln G}{d\ln\xi}+2\right)+\gamma(V-1)\frac{d V}{d\ln\xi}
=\gamma V(\frac{5}{4}-V)-\frac{25}{72}\gamma M,
\label{eq27a}
\ee
\be
\frac{d V}{d\ln\xi}-(1-V)\frac{d\ln G}{d\ln\xi}=-3V+\frac{5}{2},
\label{eq28a}
\ee
\be
\frac{d\ln Z}{d\ln\xi}-(\gamma-1)\frac{d\ln G}{d\ln\xi}=-\frac{5-2V-\frac{5}{2}\gamma}{1-V},
\label{eq29a}
\ee
\be
\xi\,\frac{d M}{d\xi}=3(G-M).
\label{eq29b}
\ee
The relations used here are
\be
\frac{\partial\xi}{\partial t}\bigg|_r=-\frac{4\xi}{5t},\quad  \frac{\partial\xi}{\partial r}\bigg|_t=\frac{\xi}{r}.
\label{eq30a}
\ee
\medskip
\\
A constant $\beta$ in the definition of the non-dimensional radius $\xi$ in (\ref{eq25a}) is obtained from the
explosion energy integral $E$. Due to zero energy (kinetic + gravitational) in the non-perturbed solution, the conserving value of the explosion energy behind the shock, in
the uniformly expanding medium, with velocity and density distributions (\ref{eq20a}), with account of the gravitational energy, is determined as
\be
\label{eq52a}
E=\int_0^{R(t)} \rho\left[\frac{v^2}{2}+\frac{c^2}{\gamma(\gamma-1)}\right]4\pi r^2 dr-
\int_0^{R(t)}\frac{G_g m dm}{r}.
\ee
In non-dimensional variables (\ref{eq24a}) this relation reduces to the equation for the constant $\beta$
\be
\label{eq53a}
\beta^{-5}=\frac{64\pi}{25}\int_0^1 G\left[\frac{V^2}{2}+\frac{Z}{\gamma(\gamma-1)}\right]\xi^4 d\xi-\frac{8}{3}\int_0^1 G\xi\left(\int_0^\xi G\eta^2 d\eta\right)d\xi.
\ee

\section{Approximate analytic solution}

\subsection{Approximate first integral}

 Using the procedure described in \cite{llhydro} for the case of the shock in a static media, it was possible to obtain an approximate energy conservation integral in the expanding medium of the universe \cite{bk15}, in the form

\be
Z=\frac{(\gamma-1)(1-V)(V-\frac{5}{6})^2}{2(V-\frac{5}{6}-\frac{1}{6\gamma})}.
\label{eq34a}
\ee
At the shock  $r=R$, $\xi=1$, using $Z(1)$ and $V(1)$ from (\ref{eq26a}), the approximate first integral gives an identity.
Using (\ref{eq34a}) we may consider only two differential equations
(\ref{eq28a}) and (\ref{eq29a}), for finding an analytical solution of the problem, similar to the classical Sedov case. The relation (\ref{eq34a}) may be interpreted as a happy choice of the profiling function for the temperature distribution behind the shock.

\subsection{Approximate analytic solution for expanding medium}

Excluding $Z$ from equations (\ref{eq28a}),(\ref{eq29a}) with the help of (\ref{eq34a}), the analytic solution of self-similar system of equations (\ref{eq27a})-(\ref{eq29b}) was obtained in \cite{bk15,bkp} in the form

\be
\label{eq40a}
\left[(\gamma+1)(3V-\frac{5}{2})\right]^{\mu_1}
\left[\frac{\gamma+1}{\gamma-1}(6\gamma V-5\gamma-1)\right]^{\mu_2}
\left[6(\gamma+1)\frac{3\gamma V-V-\frac{5}{2}}{15\gamma^2+\gamma-22}\right]^{\mu_3}=\xi,
\ee
with
\be
\mu_1= \frac{2}{15\gamma-20},\,\,\, \mu_2=\frac{\gamma-1}{17\gamma-15\gamma^2+1},
\label{eq41a}
\ee
$$\mu_3=-\frac{\gamma+1}{3\gamma-1}-\frac{\gamma-1}{17\gamma-15\gamma^2+1}
+\frac{2}{20-15\gamma}.$$

\be
\label{eq51a}
G(V)=\frac{\gamma+1}{\gamma-1}\left[6\frac{(\gamma+1)(1-V)}{\gamma-1}\right]^{\kappa_1}
\left[\frac{\gamma+1}{\gamma-1}(6\gamma V-5\gamma-1)\right]^{\kappa_2}
\ee
$$
\times
\left[\frac{3(\gamma+1)}{15\gamma^2+\gamma-22}[(6\gamma-2)V-5)]\right]^{\kappa_3}.
$$
Here
$$
\kappa_1=\frac{7}{3\gamma-1}-\frac{2}{6\gamma-7}+\frac{(15\gamma-20)(\gamma-1)}{(6\gamma-7)(15\gamma^2-17\gamma-1)}
$$
\be
\label{eq50a}
-\frac{3\gamma(15\gamma-20)}{(3\gamma-1)(15\gamma^2-17\gamma-1)}-\frac{15\gamma-20}{3\gamma-1}\,\frac{\gamma+1}{6\gamma-7},
\ee
$$
\kappa_2=-\frac{3}{3\gamma-1}+\frac{3\gamma(15\gamma-20)}{(3\gamma-1)(15\gamma^2-17\gamma-1)}.
$$
$$
\kappa_3=\frac{2}{6\gamma-7}-\frac{(15\gamma-20)(\gamma-1)}{(6\gamma-7)(15\gamma^2-17\gamma-1)}
+\frac{15\gamma-20}{3\gamma-1}\,\frac{\gamma+1}{6\gamma-7},
$$
The function $Z(V)$ is determined by the integral (\ref{eq34a}).
Here the boundary conditions (\ref{eq26a}) at $\xi=1$ have been used.

\be
M(\xi)=3\,\xi^{-3}\,\int_0^\xi G(\eta)\eta^2 d\eta.
\label{eq52b}
\ee

\section{Main properties of the approximate analy\-tic so\-lu\-ti\-on}

\subsection{Approximate analytic solution at $\gamma$ less than critical value}

\begin{figure}
\center{\includegraphics[width=1.0\linewidth]{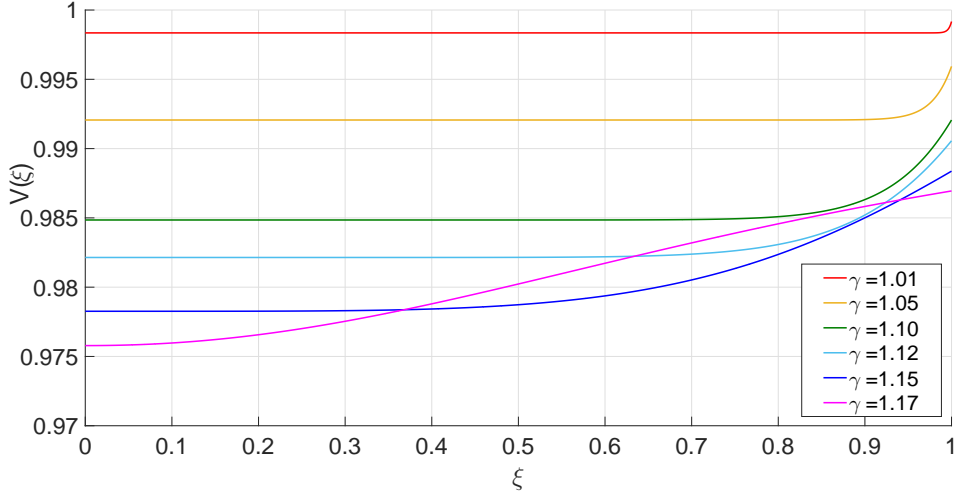}}
\caption{Approximate analytic solution without voids for $V(\xi)$}
\label{im1}
\end{figure}

\begin{figure}
\center{\includegraphics[width=1.0\linewidth]{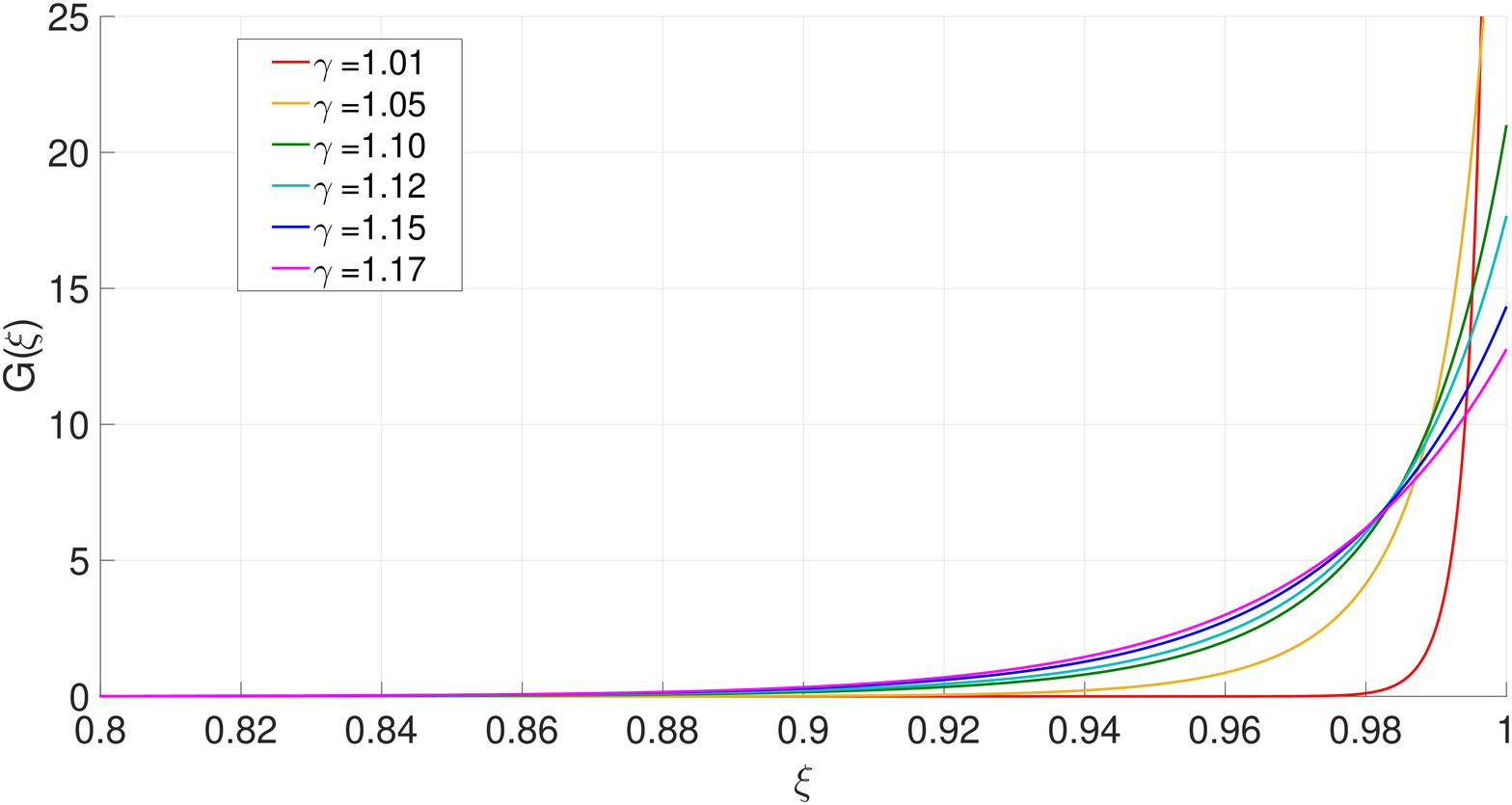}}
\caption{Approximate analytic solution without voids for $G(\xi)$}
\label{im2}
\end{figure}

\begin{figure}
\center{\includegraphics[width=1.0\linewidth]{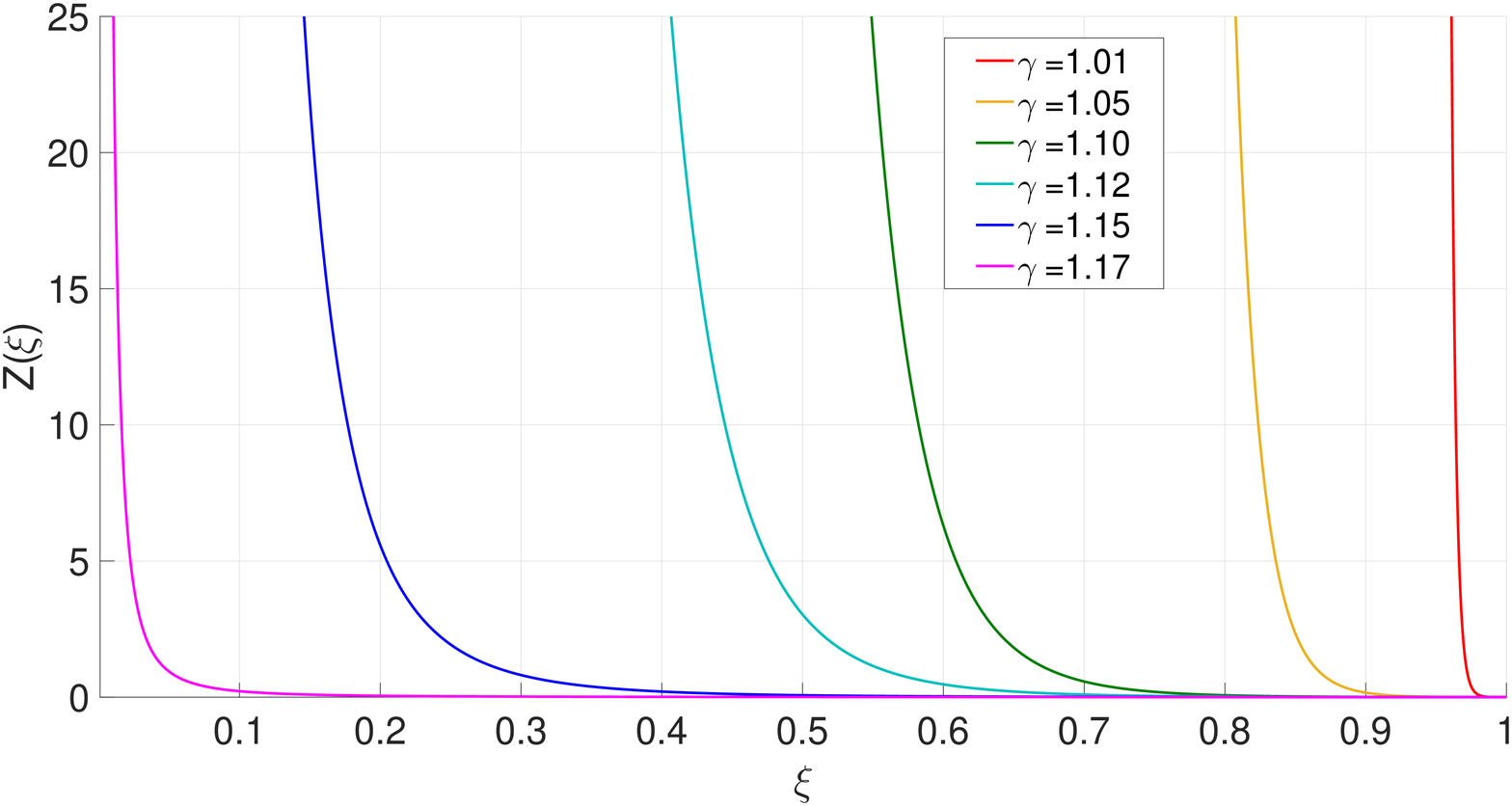}}
\caption{Approximate analytic solution without voids for $Z(\xi)$}
\label{im3}
\end{figure}

\begin{figure}
\center{\includegraphics[width=1.0\linewidth]{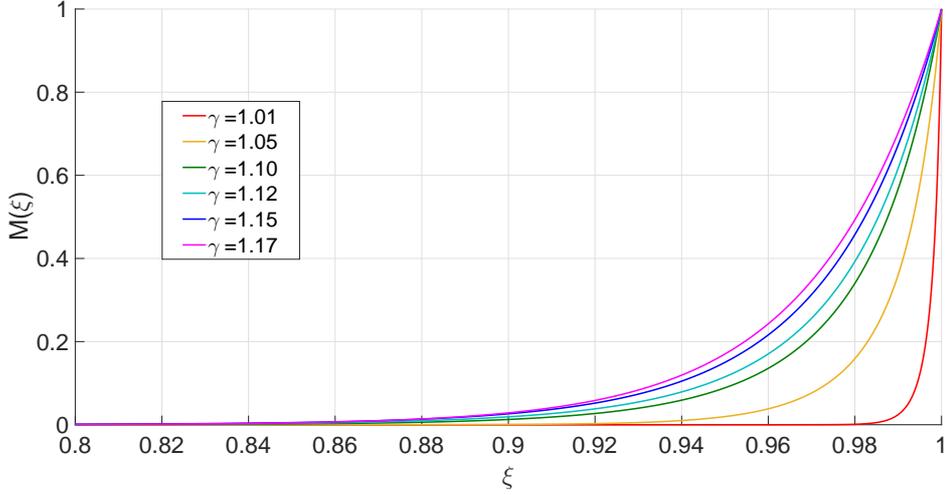}}
\caption{Solution without voids for $M(\xi)$ from (\ref{eq52b}) based on approximate analytic equations}
\label{im4}
\end{figure}

The analytic solution  (\ref{eq40a}),(\ref{eq51a}),(\ref{eq34a}),(\ref{eq52b}) has a complicated dependence of $\gamma$, and physically relevant solution exists only for limited values on $\gamma$. In order to have positive values in brackets of (\ref{eq40a}), and to satisfy the condition  for $V$ on the shock (\ref{eq26a}) we obtain restrictions for $V$ as
\be
V>\frac{5}{6},\quad V>\frac{1+5\gamma}{6\gamma}, \quad V<V(1)=\frac{5\gamma+7}{6(\gamma+1)}.
\ee
To satisfy all these conditions we obtain the restriction for $\gamma$ as $1<\gamma<\gamma_*$, where
$\gamma_*$ is defined by equation
\be
\label{eq53b}
15\gamma^2+\gamma-22=0, \qquad \gamma_*=-\frac{1}{30}+\sqrt{\frac{1}{900}+\frac{22}{15}},\qquad
 \gamma_*\approx 1.1782.
\ee
Numerical solutions of self-similar equations (\ref{eq27a})-(\ref{eq29b}), presented below, have similar restrictions for $\gamma$. We may conclude, therefore, that for other $\gamma>\sim\gamma_*$ there are no smooth self-similar solutions in the whole space. On figures are plotted, for different $\gamma<\gamma_*$,  functions from the analytical solution: $V(\xi)$  from (\ref{eq40a}) in Fig.\ref{im1}; $G(\xi)$ from (\ref{eq51a}) in Fig.\ref{im2}; $Z(\xi)$ from (\ref{eq34a}) in Fig.\ref{im3}; and $M(\xi)$ from (\ref{eq52b}) in Fig.\ref{im4}.

Introduce notations

\be
\label{eq54b}
 V'=\frac{d\,V}{d\,\xi},\quad
 G'=\frac{d\,G}{d\,\xi},\quad
 Z'=\frac{d\,Z}{d\,\xi}
\ee
At the shock $\xi=1$ the derivative of the self-similar functions are found from the analytic solution (\ref{eq40a})-(\ref{eq51a}) in the form \cite{bkp}

\begin{eqnarray}
 V'(1)=\frac{-15\gamma^2-\gamma+22}{6(\gamma+1)^2};\quad
 G'(1)= \frac{-15\gamma^2+5\gamma+28}{(\gamma-1)^2};\nonumber\\
\quad Z'(1)= \frac{(15\gamma^2+\gamma-22)\gamma}{9(\gamma+1)^3}.\qquad\qquad
\label{eq54cc}
\end{eqnarray}
It follows from (\ref{eq53b}),(\ref{eq54cc}), that for $\gamma<\gamma_*$ the derivatives have the following signs
\be
 V'(1)>0;\quad G'(1)>0;\quad  Z'(1)<0
\label{eq54d}
\ee

\subsection{Approximate analytic solution at $\gamma$ larger than critical value}

Consider approximate analytic solution at $\gamma \geq \gamma_*\approx 1.1782$. Contrary to the approximate analytic solution for $V(\xi)$ at $\gamma < \gamma_*$, the function $V(\xi)$ increases up to infinity at $\xi \rightarrow 0$.

\begin{figure}
\center{\includegraphics[width=0.9\linewidth]{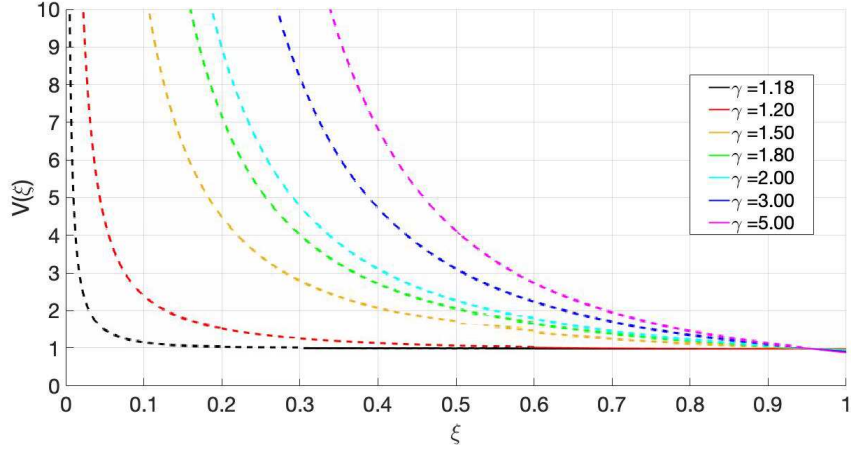}}
\caption{Approximate analytic solution for $V(\xi)$ at $\gamma > \gamma_*$, plotted according to Eq.(\ref{eq40a}). Non-physical parts of curves at $V\geq 1$ are given by dashed lines.  }
\label{im5f}
\end{figure}

\begin{figure}
\center{\includegraphics[width=0.9\linewidth]{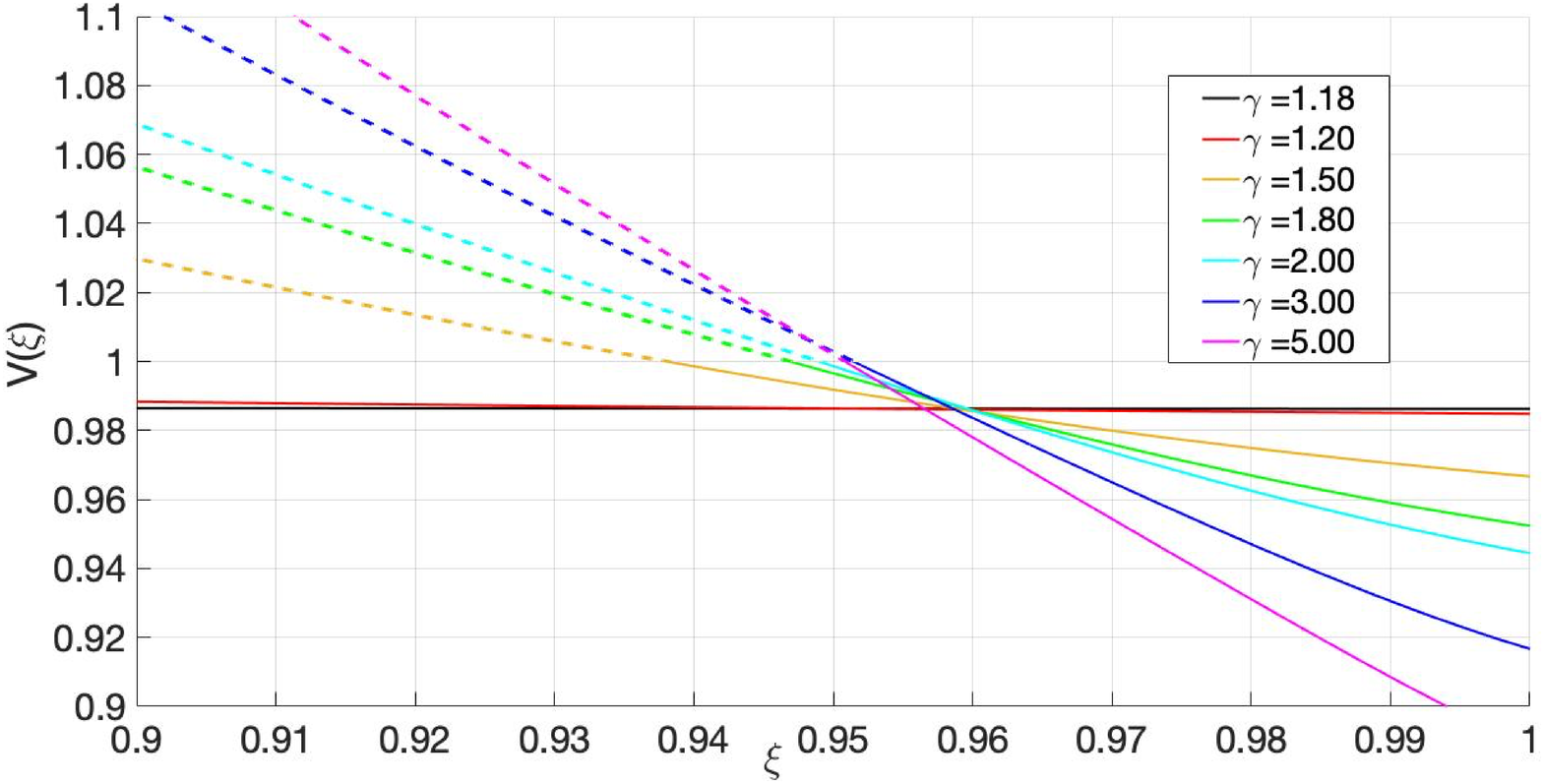}}
\caption{Approximate analytic solution for $V(\xi)$ at $\gamma > \gamma_*$, plotted according to Eq.(\ref{eq40a}) in the vicinity of the shock. Non-physical parts of curves at $V\geq 1$ are given by dashed lines.}
\label{im6}
\end{figure}

\begin{figure}
\center{\includegraphics[width=1.0\linewidth]{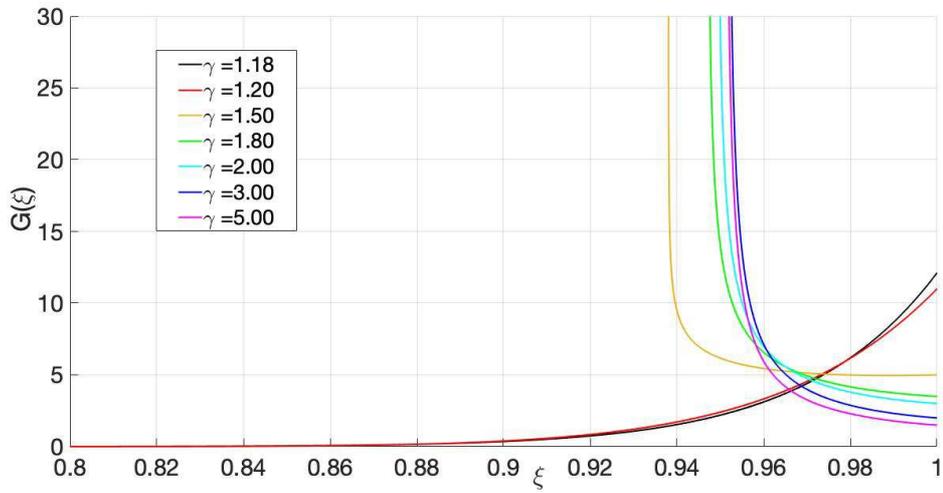}}
\caption{Approximate analytic solution for $G(\xi)$ at $\gamma > \gamma_*$, plotted according to Eqs.(\ref{eq40a}),(\ref{eq51a}).}
\label{im7}
\end{figure}

\begin{figure}
\center{\includegraphics[width=1.0\linewidth]{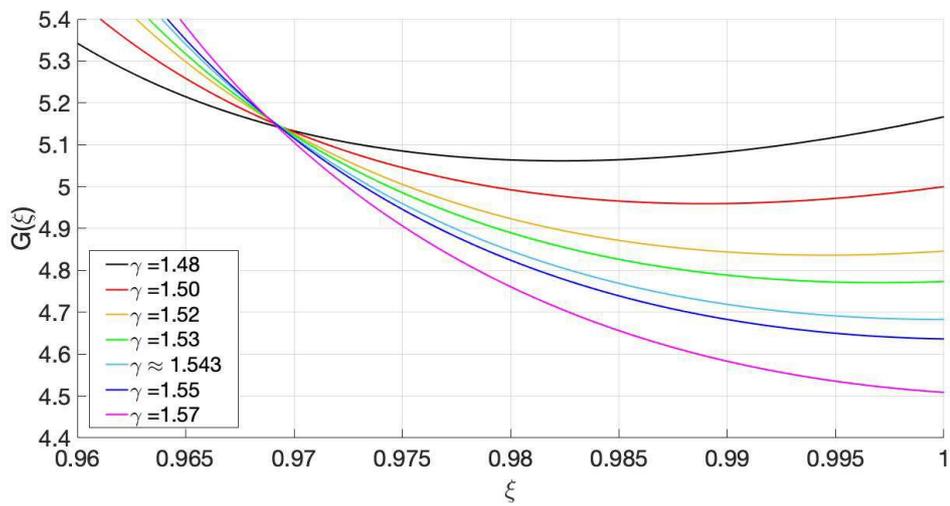}}
\caption{Approximate analytic solution for $G(\xi)$ at $\gamma \approx 1.1543 $, in the vicinity of the shock.}
\label{im8}
\end{figure}

\begin{figure}
\center{\includegraphics[width=1.0\linewidth]{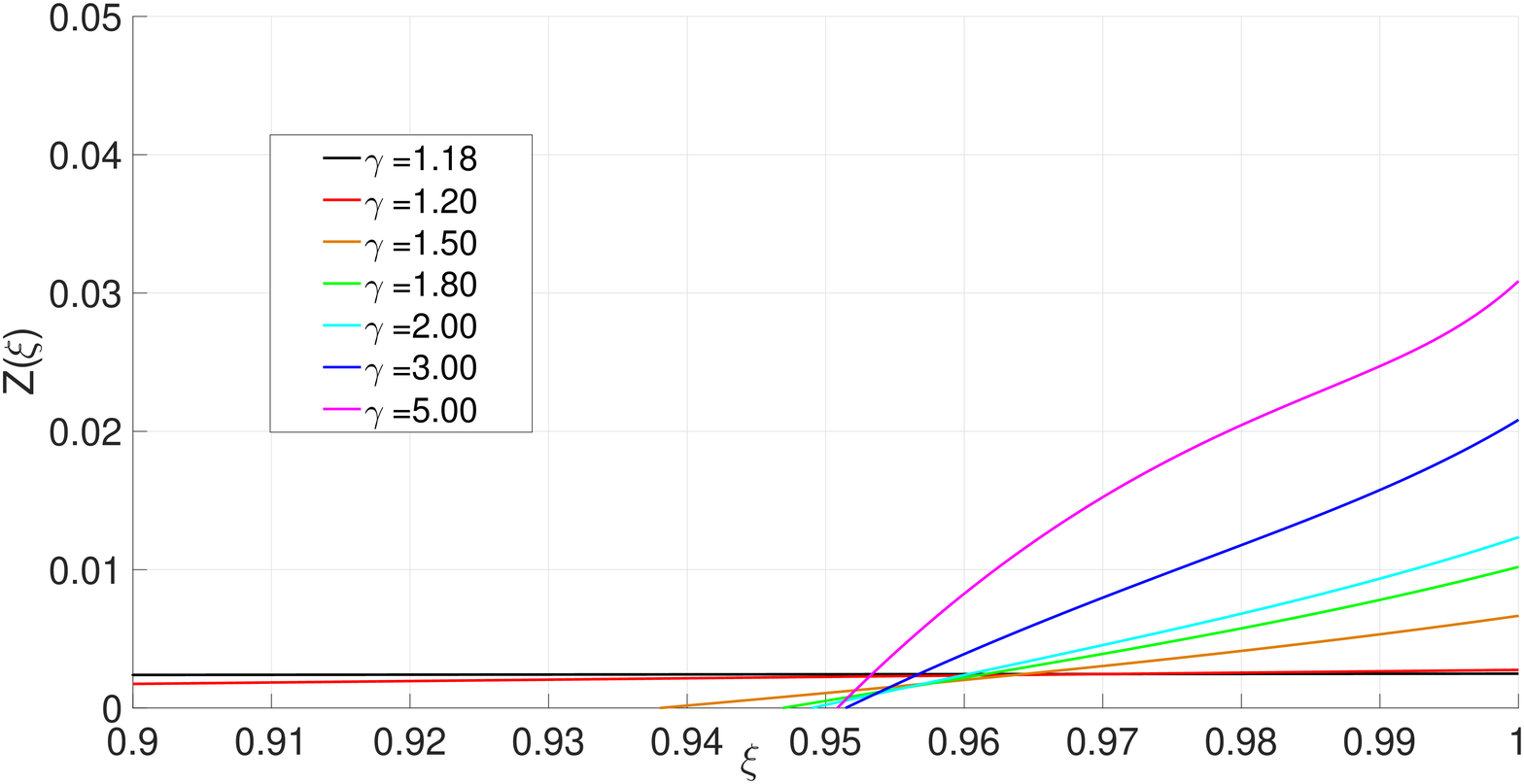}}
\caption{Approximate analytic solution for $Z(\xi)$ at $\gamma >\gamma_*$} plotted according to Eq.(\ref{eq40a}),(\ref{eq34a}) in the vicinity of the shock.
\label{im9}
\end{figure}

\begin{figure}
\center{\includegraphics[width=1.0\linewidth]{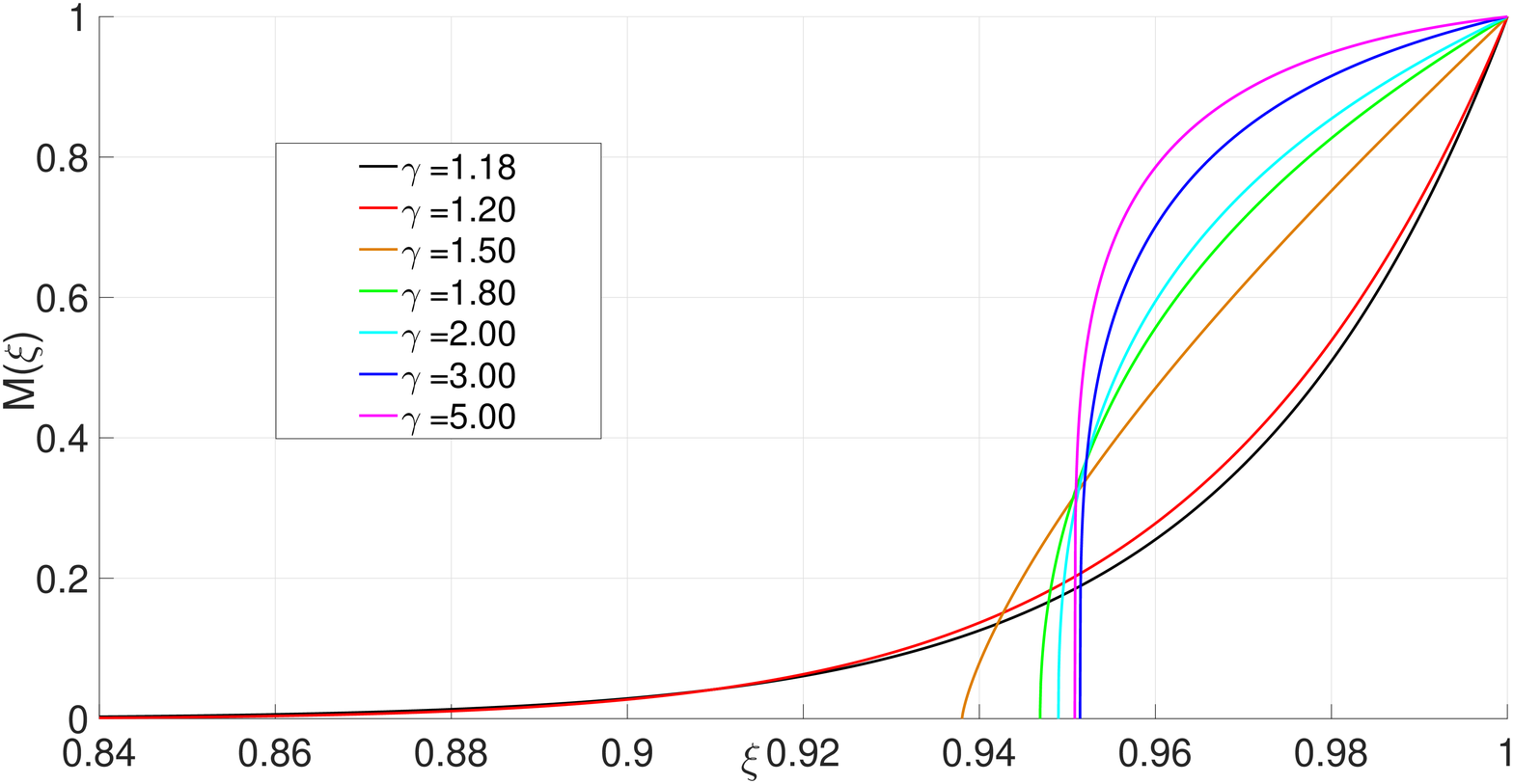}}
\caption{Approximate analytic solution for $M(\xi)$ at $\gamma > \gamma_*$, plotted by integration in Eq.(\ref{eq52b}) in the vicinity of the shock.}
\label{im10}
\end{figure}

\begin{figure}
\center{\includegraphics[width=1.0\textwidth]{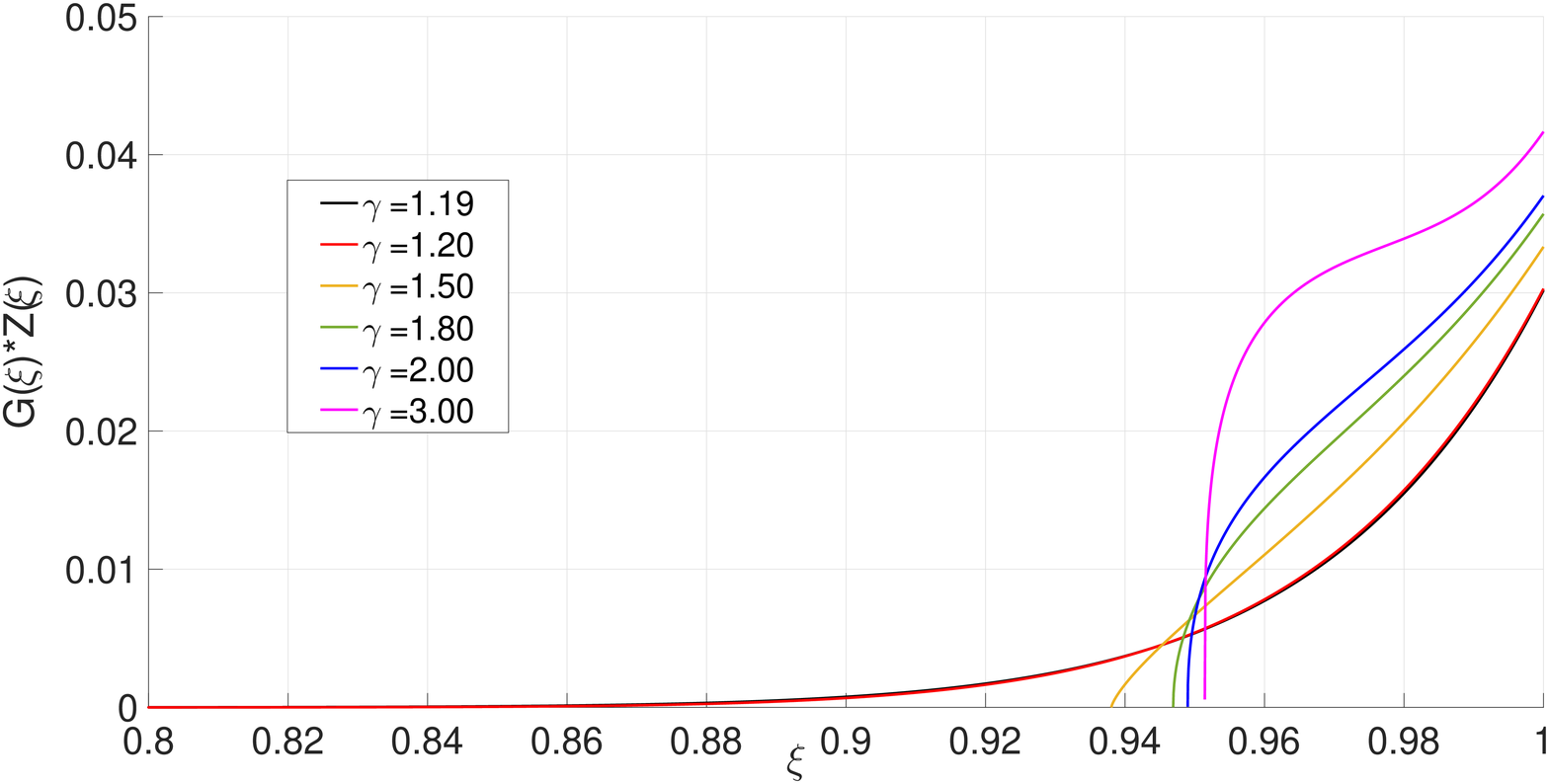}}
\caption{Approximate analytic solution for $G(\xi)*Z(\xi)$ at big $\gamma$, in the vicinity of the shock.}
\label{im101}
\end{figure}

It follows from (\ref{eq51a}) that $G(\xi)$ has a physical sense only when $V(\xi)<1$, because $V(\xi)=1$ is the point where $G(\xi)=0$. That means that there is a point where density of matter becomes zero and spherical void area appears. Dependence of radius $\xi$ of such spherical void areas on $\gamma$ can be written in the form
\be
\left[\frac{\gamma+1}{2}\right]^{\mu_1}
\bigg[\gamma+1\bigg]^{\mu_2}
\left[3(\gamma+1)\frac{6\gamma -7}{15\gamma^2+\gamma-22}\right]^{\mu_3}=\xi,
\ee
with $\mu_1,\,\,\mu_2,\,\,\mu_3$ from Eq.(\ref{eq41a}).

Calculation of self-similar variables, using Eqs. (\ref{eq40a}),(\ref{eq51a}) gives, that at the point
with $V=1$ the density goes to zero at $\gamma<\gamma_{cr1}=1.4$, and for larger $\gamma$ the density tends to infinity at this point. Nevertheless, the temperature goes to zero at this point, so that the pressure, represented by the function $GZ$ goes to zero at the inner edge of the layer
at $V=1$, so we obtain a self-consistent solution with the spherical void.
The following figures  represent behaviour of functions at different $\gamma>\gamma_*$:  $V(\xi)$ in Figs.(\ref{im5f}),(\ref{im6}); $G(\xi)$ in Figs.(\ref{im7}),(\ref{im8}); $Z(\xi)$ in Fig.(\ref{im9});
$M(\xi)$ in Fig.(\ref{im10}); $G(\xi)\times Z(\xi)$ in Fig.(\ref{im101}).

We obtain from (\ref{eq54cc}) that
$G'(\xi)|_{\xi=1}>0$ at $\gamma<\frac{5+\sqrt{1705}}{30}\approx 1.54305$ and $G'(\xi)|_{\xi=1}<0$ at $\gamma>\frac{5+\sqrt{1705}}{30}. $
So the density starts to fall and then rises up to infinity at
$1.4<\gamma <\frac{5+\sqrt{1705}}{30}$. When $\gamma >\gamma_2=\frac{5+\sqrt{1705}}{30}$
the density starts to grow inside from the shock, and continues rising up to infinity.

\section{Numerical solution of self-similar equations}

\subsection{Numerical solution at $\gamma$ less than critical value}

\begin{figure}
\center{\includegraphics[width=1\linewidth]{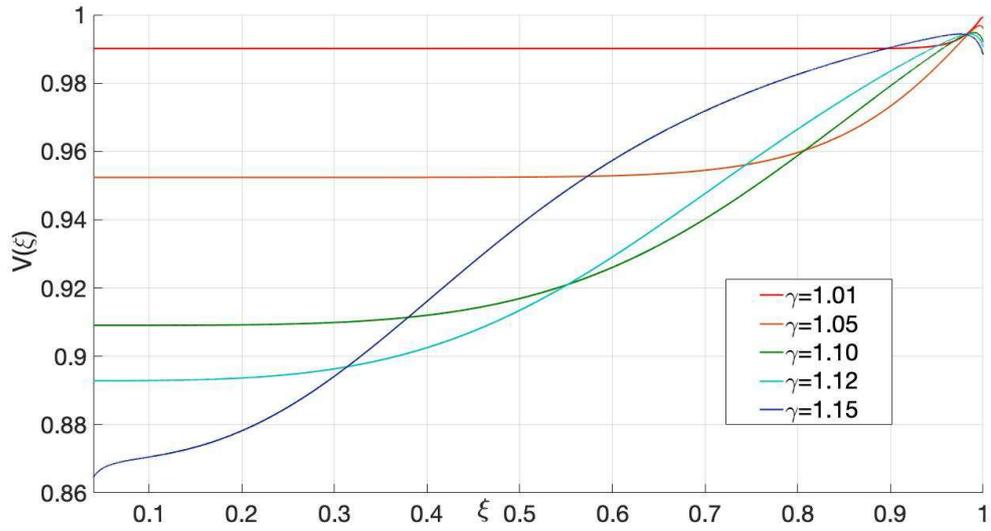}}
\caption{Numerical solution for $V(\xi)$.}
\label{im11}
\end{figure}

\begin{figure}
\center{\includegraphics[width=1.0\linewidth]{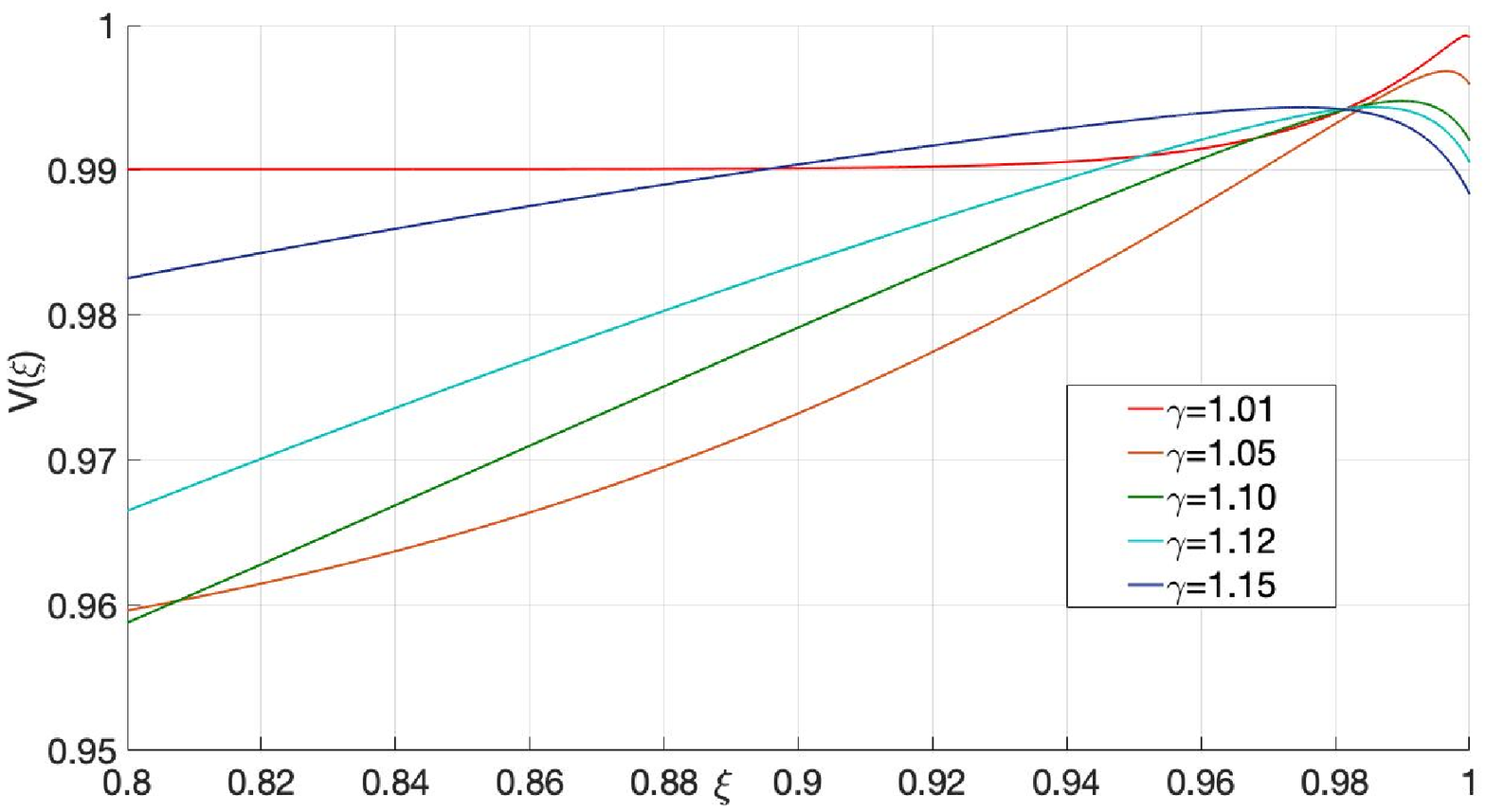}}
\caption{Numerical solution for $V(\xi)$ at $\xi$ from 0.8 to 1.0.}
\label{im12}
\end{figure}

\begin{figure}
\center{\includegraphics[width=1.0\linewidth]{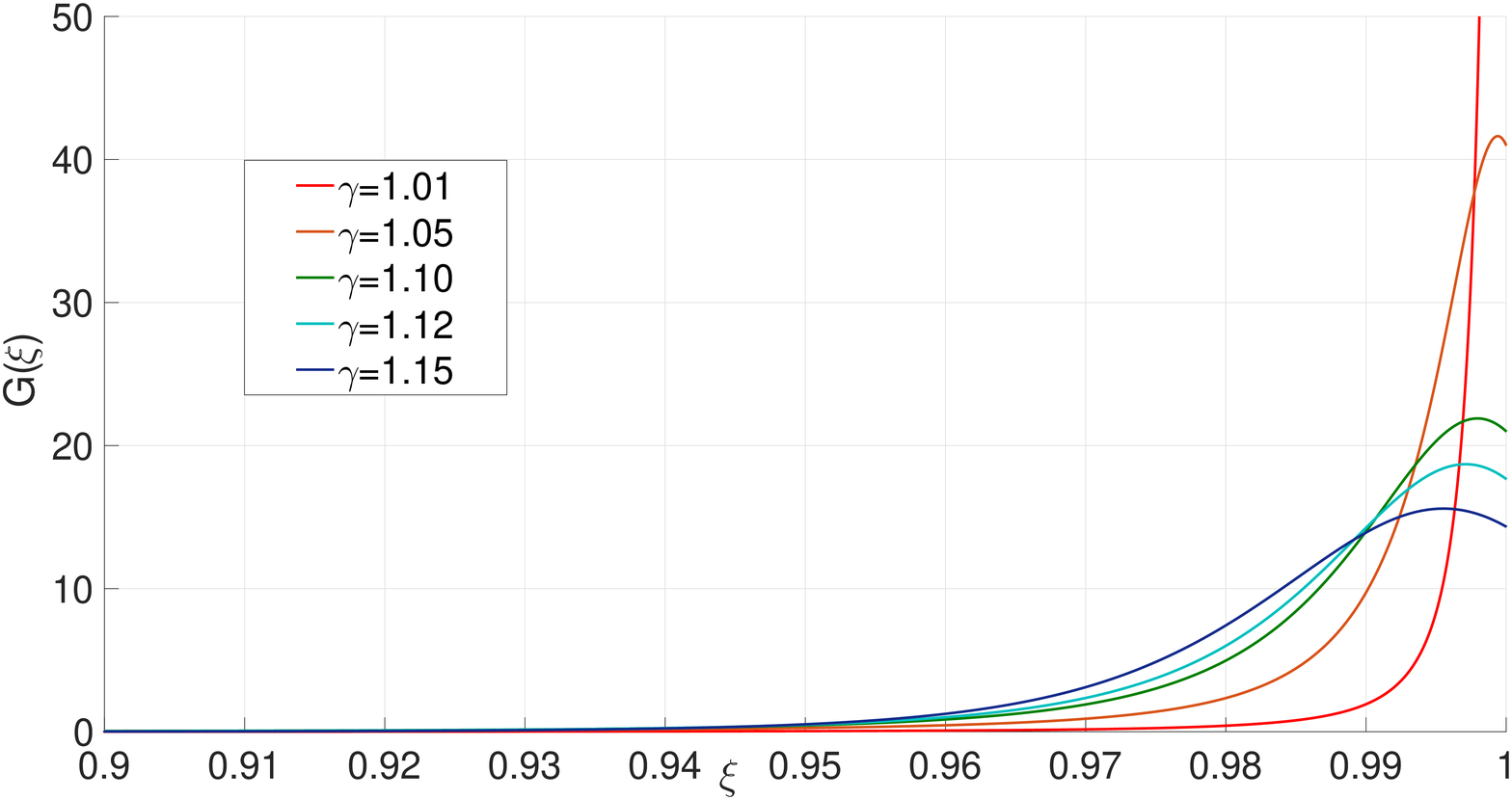}}
\caption{Numerical solution for $G(\xi)$ at $\xi$ from 0.9 to 1.0.}
\label{im13}
\end{figure}

\begin{figure}
\center{\includegraphics[width=1.0\linewidth]{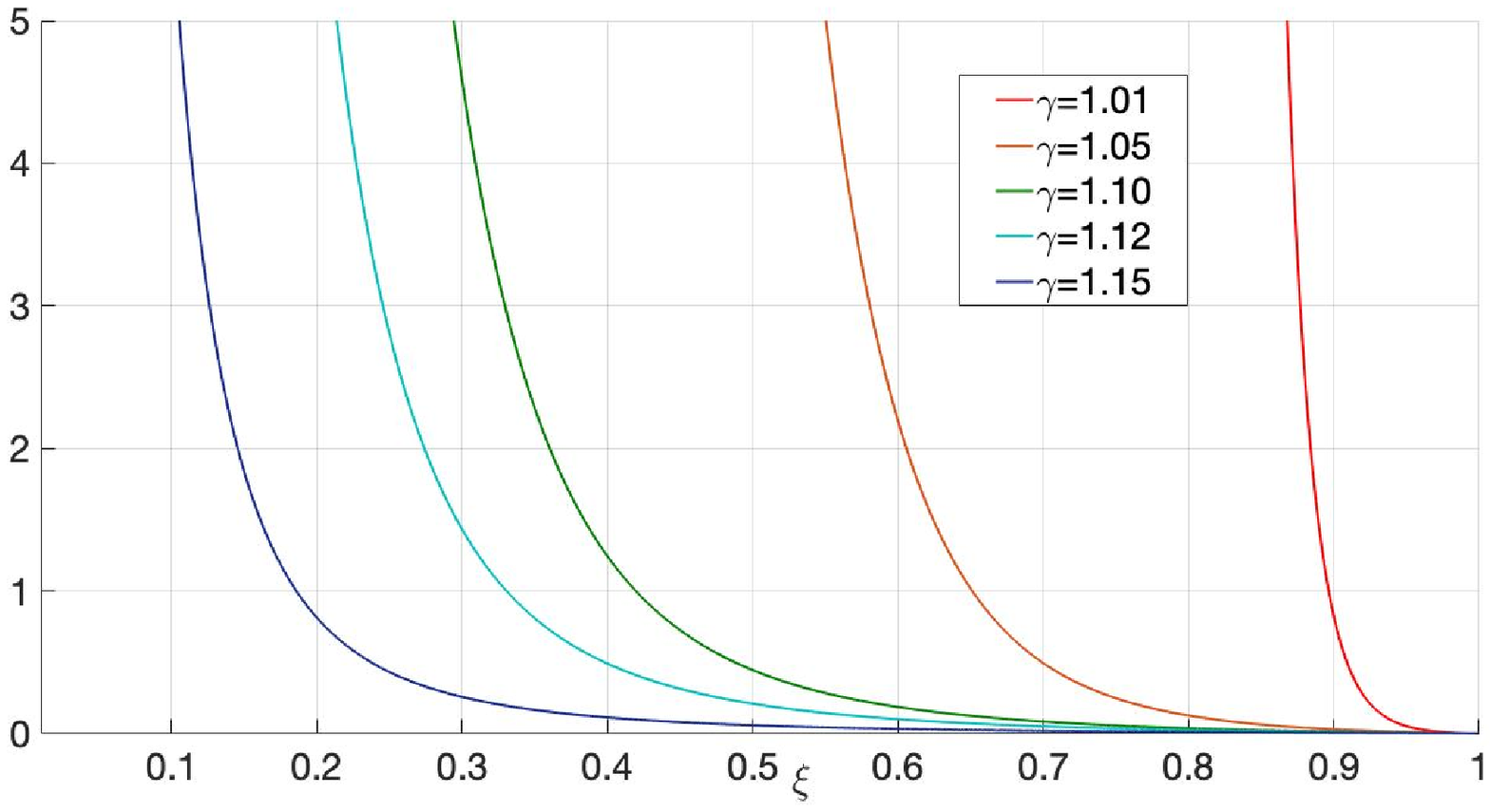}}
\caption{Numerical solution for $Z(\xi)$.}
\label{im14}
\end{figure}

\begin{figure}
\center{\includegraphics[width=1.0\linewidth]{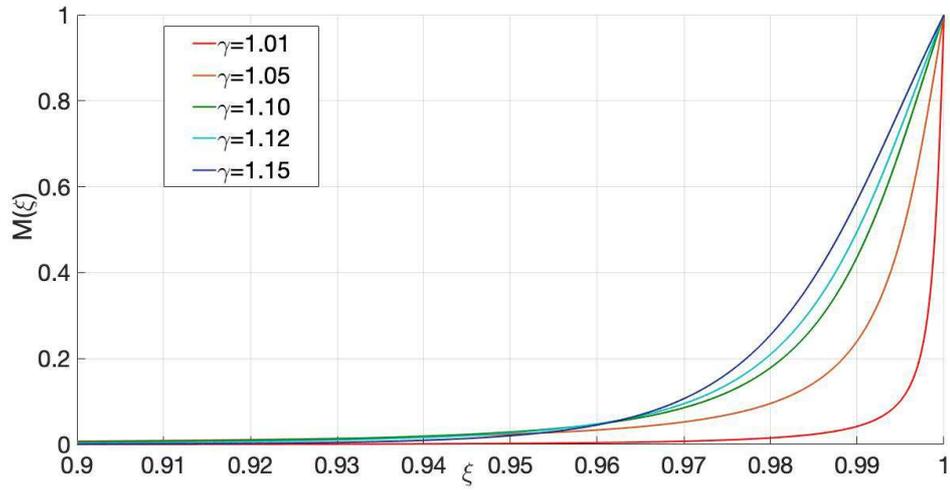}}
\caption{Numerical solution for $M(\xi)$ at $\xi$ from 0.9 to 1.0.}
\label{im15}
\end{figure}

The system of equations (\ref{eq27a})-(\ref{eq29b}) written explicitly for derivatives has a form:  \\
\begin{equation*}
\label{eq54c}
\begin{cases}
$$\frac{dlnG}{dln\xi}=\frac{\frac{3-\frac{5}{2}\gamma}{1-V}Z-\frac{25}{72}\gamma M   +\gamma(2V^2-\frac{17}{4}V+\frac{5}{2})}{\gamma[Z-(1-V)^2]};$$\\
$$\frac{dV}{dln\xi}=(1-V)\frac{dlnG}{dln\xi}-3V+\frac{5}{2};$$\\
$$\frac{dlnZ}{dln\xi}=(\gamma-1)\frac{dlnG}{dln\xi}-\frac{5-2V-\frac{5}{2}\gamma}{1-V};$$\\
$$\frac{dM}{dln\xi}=3(G-M)$$ \\
\end{cases}
\end{equation*}
That reduces to:
\be
\label{eq55b}
\xi\frac{dG}{d\xi}=G\frac{\frac{3Z}{\gamma}\frac{1-\frac{5\gamma}{6}}{1-V}-\frac{17}{4}V+\frac{5}{2}+2\,V^2
-\frac{25}{72}M}{Z
-(1-V)^2},\quad \xi\,\frac{d M}{d\xi}=3(G-M),
\ee
\be
\nonumber
\xi\frac{dV}{d\xi}=\xi\frac{1-V}{G}\frac{dG}{d\xi}-3(V-\frac{5}{6}),\quad
\frac{\xi}{Z}\frac{dZ}{d\xi}=\xi\frac{\gamma-1}{G}\frac{dG}{d\xi}-\frac{5-2V-\frac{5}{2}\gamma}{1-V}.
\ee
Let us note that the expression (\ref{eq52b}) for $M(\xi)$ is also valid for the exact numerical solution.
 This system is solved numerically, starting from the point $\xi=1$, where the variables are found from the conditions at the shock (\ref{eq26a}), as

\be
\label{eq57b}
\quad \frac{dV}{d\xi}\bigg|_{\xi=1}=\frac{-30\gamma^2-11\gamma+27}{6(\gamma+1)^2};\quad \frac{dG}{d\xi}\bigg|_{\xi=1}=\frac{-30\gamma^2-5\gamma+33}{(\gamma-1)^2};
\ee
\be
\nonumber
\frac{dZ}{d\xi}\bigg|_{\xi=1}= -\frac{\gamma(15\gamma^3-35\gamma^2-17\gamma+49)}{18(\gamma+1)^3};\quad
\frac{dM}{d\xi}\bigg|_{\xi=1}=\frac{6}{\gamma-1}
\ee
The sign of derivatives $V'$, $G'$ and $Z'$ is negative at $\xi=1$, what differs from the sign of some derivatives in the approximate analytic solution in (\ref{eq54d}).
It follows from the numerical integration of the system (\ref{eq55b}), that close to the shock boundary the values of $G(\xi)$ and $V(\xi)$ reach their maxima, and after decrease monotonically  until the origin $\xi=0$, see Figs.(\ref{im11})-(\ref{im13}).
 Numerical solutions for $Z(\xi)$ and $M(\xi)$ for different $\gamma$ are given in Figs.(\ref{im14})-(\ref{im15}), respectively.
      The solutions of self-similar equations without empty voids exist only in the interval
$1<\gamma<\gamma_{**}$, where $\gamma_{**}=1.155$.
 At $\gamma > \gamma_{**}=1.155$ the empty spherical void is formed
around the center, at a finite distance from the shock.
Similar voids are formed in  Sedov solution for a shock in the static uniform gas at $\gamma > 7$ \cite{llhydro}.

\newpage

\subsection{Numerical solution at $\gamma$ bigger than critical value}
Consider approximate analytic solution at $\gamma \geq \gamma_{**}\approx 1.155$. Like in approximate analytic solution, we consider radius of a spherical void  as point where velocity $V=1$. Such point is also a point where numerical solution stops its existence.

\begin{figure}
\center{\includegraphics[width=1.0\textwidth]{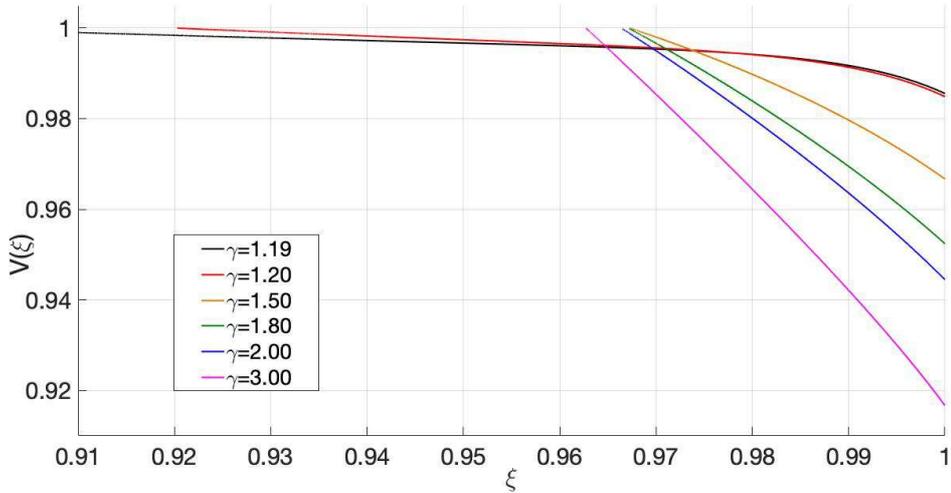}}
\caption{Numerical solution for $V(\xi)$ at big $\gamma$, at $\xi$ from 0.91 to 1.0.}
\label{im16}
\end{figure}

\begin{figure}
\center{\includegraphics[width=1.0\textwidth]{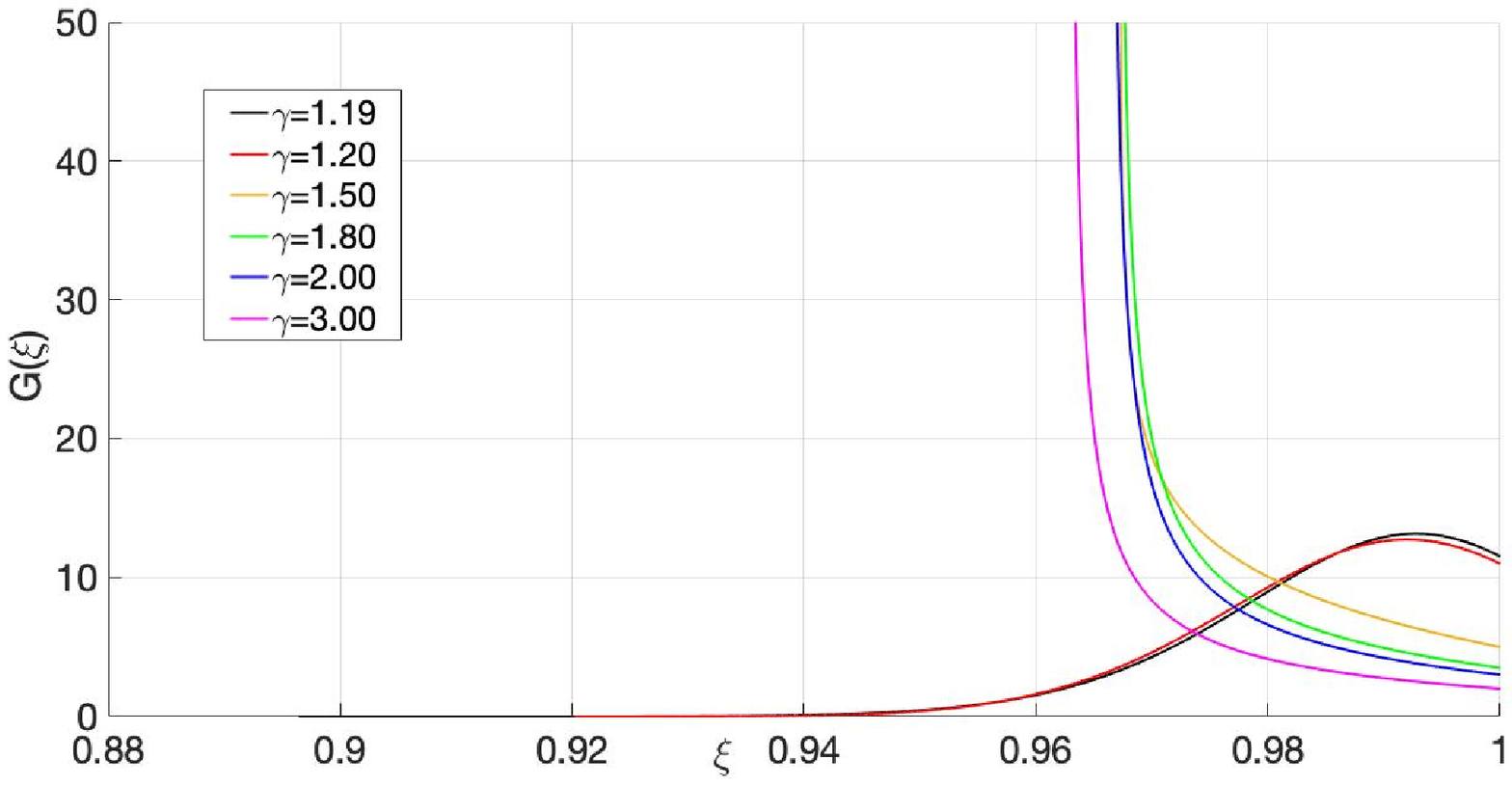}}
\caption{Numerical solution for $G(\xi)$ at big $\gamma$, at $\xi$ from 0.88 to 1.0.}
\label{im17}
\end{figure}

\begin{figure}
\center{\includegraphics[width=1.0\textwidth]{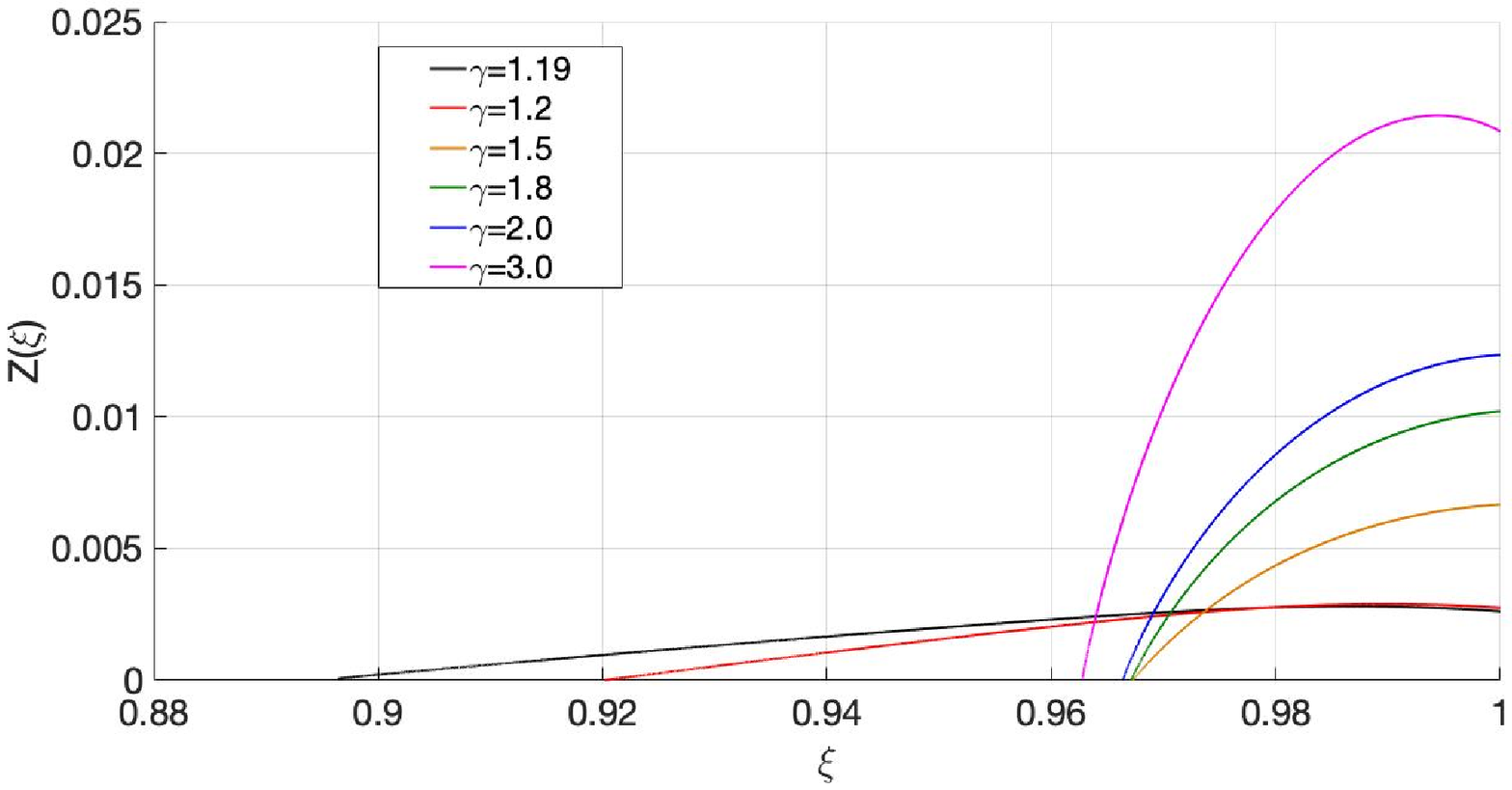}}
\caption{Numerical solution for $Z(\xi)$ at big $\gamma$, at $\xi$ from 0.88 to 1.0. }
\label{im18}
\end{figure}

\begin{figure}
\center{\includegraphics[width=1.0\textwidth]{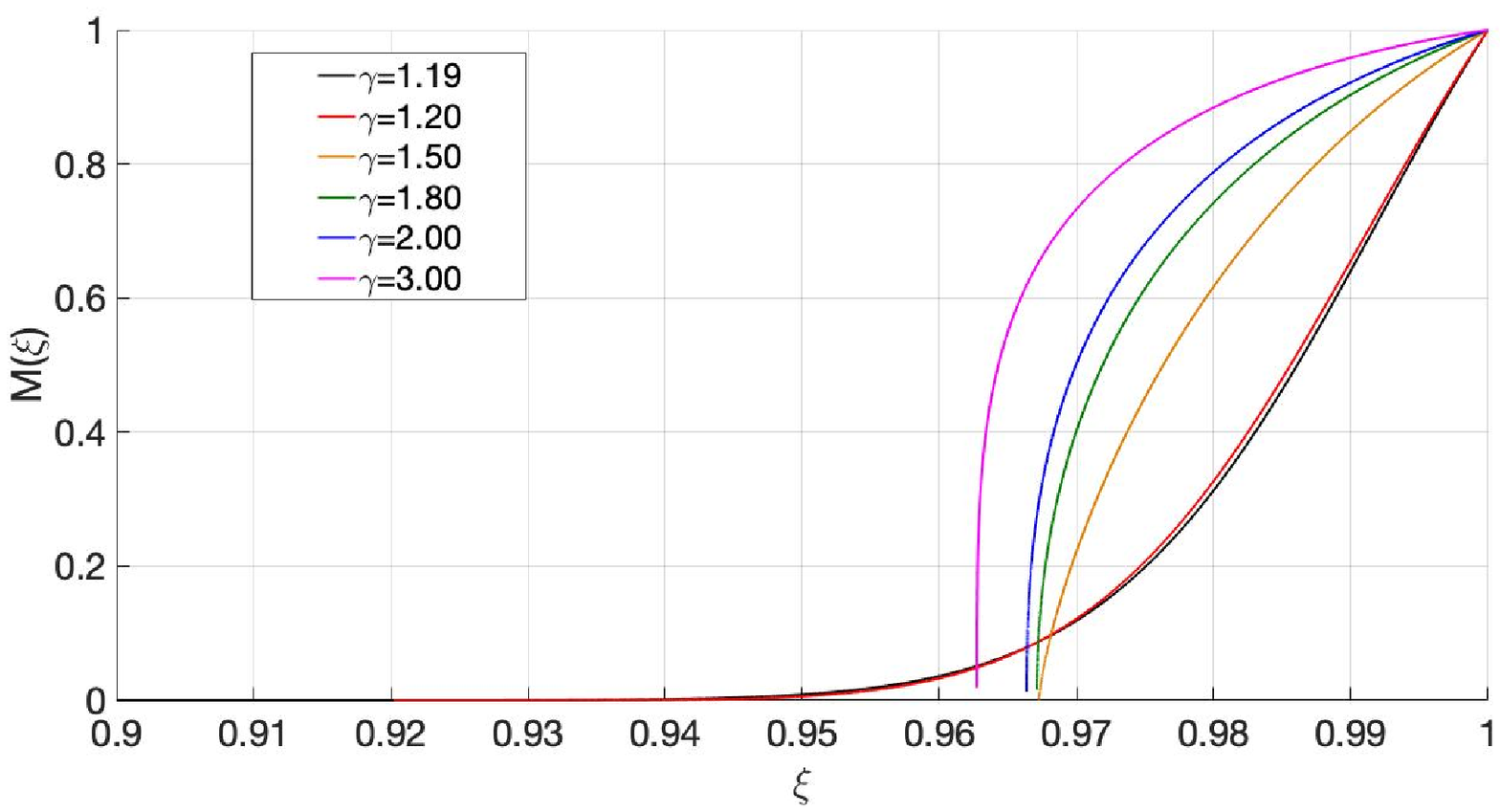}}
\caption{Numerical solution for $M(\xi)$ at big $\gamma$, at $\xi$ from 0.9 to 1.0.}
\label{im19}
\end{figure}

The important parameter is the pressure value $P \sim \rho c^2 \sim G(\xi)Z(\xi)$ at the point at $V(\xi) = 1$.

\begin{figure}
\center{\includegraphics[width=1.0\textwidth]{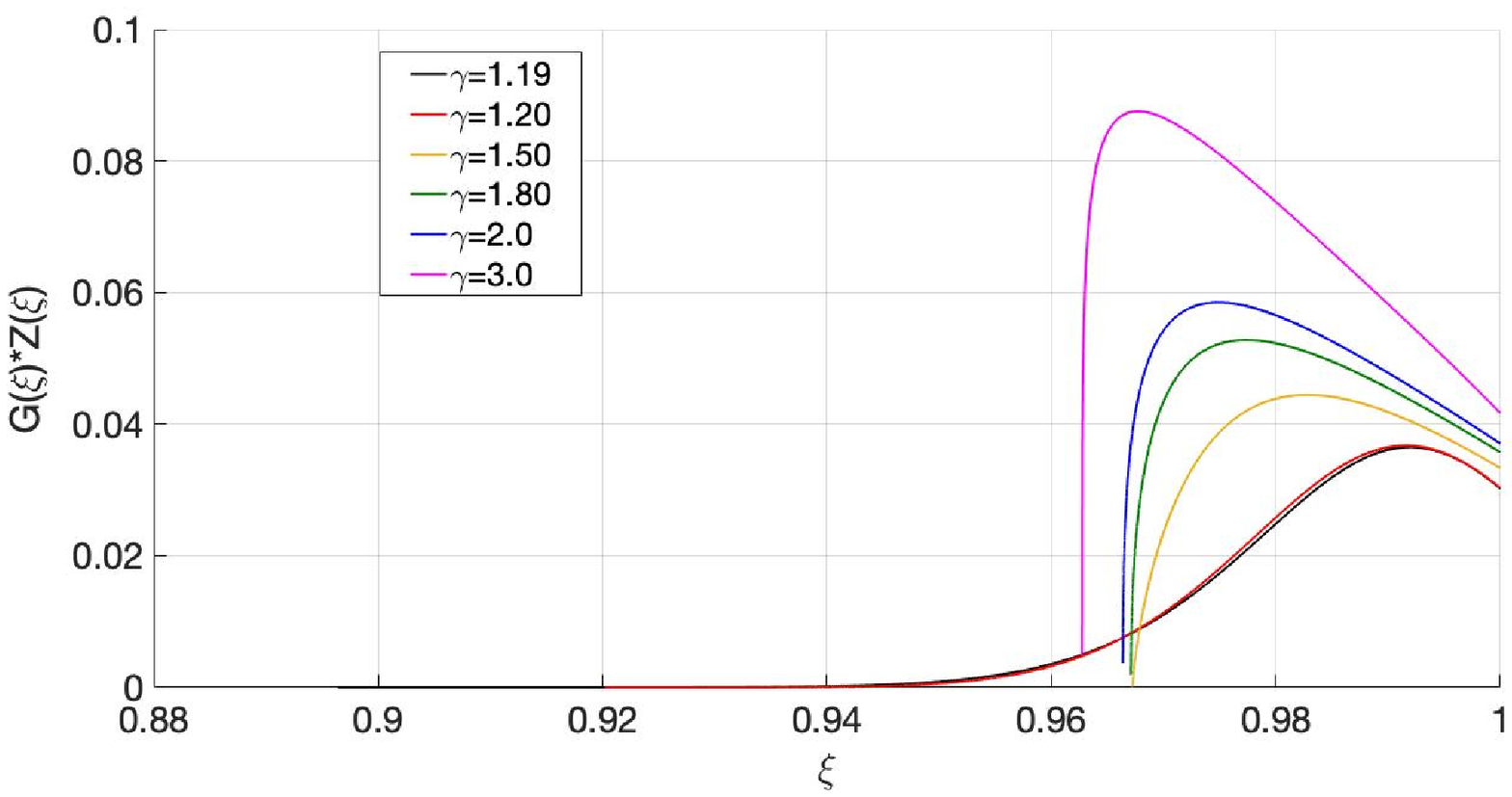}}
\caption{Numerical solution for $G(\xi)*Z(\xi)$ at big $\gamma$, at $\xi$ from 0.9 to 1.0.}
\label{im20}
\end{figure}

Calculations give that the pressure equals $0$ at $V = 1$, but the behaviour of the density $G(\xi)$ at $V=1$ depends on $\gamma$. Like in the approximate analytic solution,  at the point
with $V=1$ the density goes to zero at $\gamma<\gamma_{cr1}=1.4$, and for larger $\gamma$ the density tends to infinity at this point. Nevertheless, the temperature goes to zero at this point, so that the pressure, represented by the function $GZ$ goes to zero at the inner edge of the layer
at $V=1$. So we obtain a continuous pressure, self-consistent solution with a spherical void, with zero, or infinite density on its inner zero-pressure boundary.
The following figures  represent behaviour of functions at different $\gamma>\gamma_*$:  $V(\xi)$ in Fig.(\ref{im16}); $G(\xi)$ in Fig.(\ref{im17}); $Z(\xi)$ in Fig.(\ref{im18});
$M(\xi)$ in Fig.(\ref{im19}); $G(\xi)\times Z(\xi)$ in Fig.(\ref{im20}).

It is clear from Fig.(\ref{im20}), that on the inner boundary of the layer $P = 0$ due to zero temperature. Inside there is an empty hole. The density at the inner boundary at $\gamma>1.4$ becomes infinite instead of zero at smaller ones.

\newpage

\section{Comparison of approximate analytic and numerical so\-lu\-tions. Discussion}

 Let us compare radiuses of spherical void area in analytic $(\xi_*^{an})$ and numerical $(\xi_*^{num})$ solutions in the Table \ref{tabular:timesandtenses}.
 \begin{table}
\caption{The values $\xi_*(\gamma)$ for approximate analytic and numerical solutions}
\label{tabular:timesandtenses}
\begin{center}
\begin{tabular}{ |c |c |c|}
\hline
\textbf{$\gamma$} & \textbf{$\xi_*^{an}$} & \textbf{$\xi_*^{num}$} \\
\hline
1,18 &  0,2498 & 0,8462 \\
\hline
1,20 & 0,7364 & 0,92018 \\
\hline
1,50 & 0,938 & 0,9672 \\
\hline
2,00 & 0,94898 & 0,9664 \\
\hline
5,00 & 0,95084 & 0,9581 \\
\hline
10,00 & 0,94866 & 0,9527 \\
\hline
\end{tabular}
\end{center}
\end{table}
The analytic formula for the dependence $\xi_*^{an}$ in the analytic solution is obtained from (\ref{eq40a}) at $V=1$. We have

\be
\label{eq40aa}
\xi_*^{an}=\frac{(\gamma+1)^{\mu_1+\mu_2+\mu_3}}{2^{\mu_1}}
\left(\frac{18\gamma -21}{15\gamma^2+\gamma-22}\right)^{\mu_3},
\ee
with powers from (\ref{eq41a}) as

\begin{eqnarray}
\mu_1= \frac{2}{15\gamma-20},\qquad \mu_1+\mu_2+\mu_3=-\frac{\gamma+1}{3\gamma-1},
\label{eq41aa}
\\
\mu_3=-\frac{\gamma+1}{3\gamma-1}-\frac{\gamma-1}{17\gamma-15\gamma^2+1}
+\frac{2}{20-15\gamma} \nonumber.
\end{eqnarray}
Tending formally $\gamma \rightarrow \infty$ we obtain from (\ref{eq40aa}),(\ref{eq40aa}) the value

\be
\xi_*^{an}(\infty)=\left(\frac{5}{6}\right)^{1/3} = 0.941.
\label{42aa}
\ee
We see from the Table \ref{tabular:timesandtenses} the value of $\xi_*$ has its maximum value both in analytic and numerical models. It indicates the thickness of the layer goes through the minimum. For $\gamma=10$ the value of $\xi_*^{an}$ is close to its limiting value in (\ref{42aa}). Actually the results for large $\gamma > \sim 5$, which is obtained from self-similar solution, are not reliable. At large $\gamma$ the matter compressibility decreases, and the shock is becoming weaker. Hugoniot relations in the form (\ref{eq23a}) describing the strong shock are not valid anymore. With general Hugoniot adiabatic relations \cite{llhydro} we cannot construct a self-similar solution. Therefore the results for large $\gamma$ could be considered only as rough estimations by the order of magnitude. The maximum value of $(\xi_*^{num})$ in the Table \ref{tabular:timesandtenses} is related to the minimal thickness of the layer for large $\gamma$.

\begin{figure}[h]
\begin{minipage}[h]{0.5\linewidth}
\center{\includegraphics[width=1.0\linewidth]{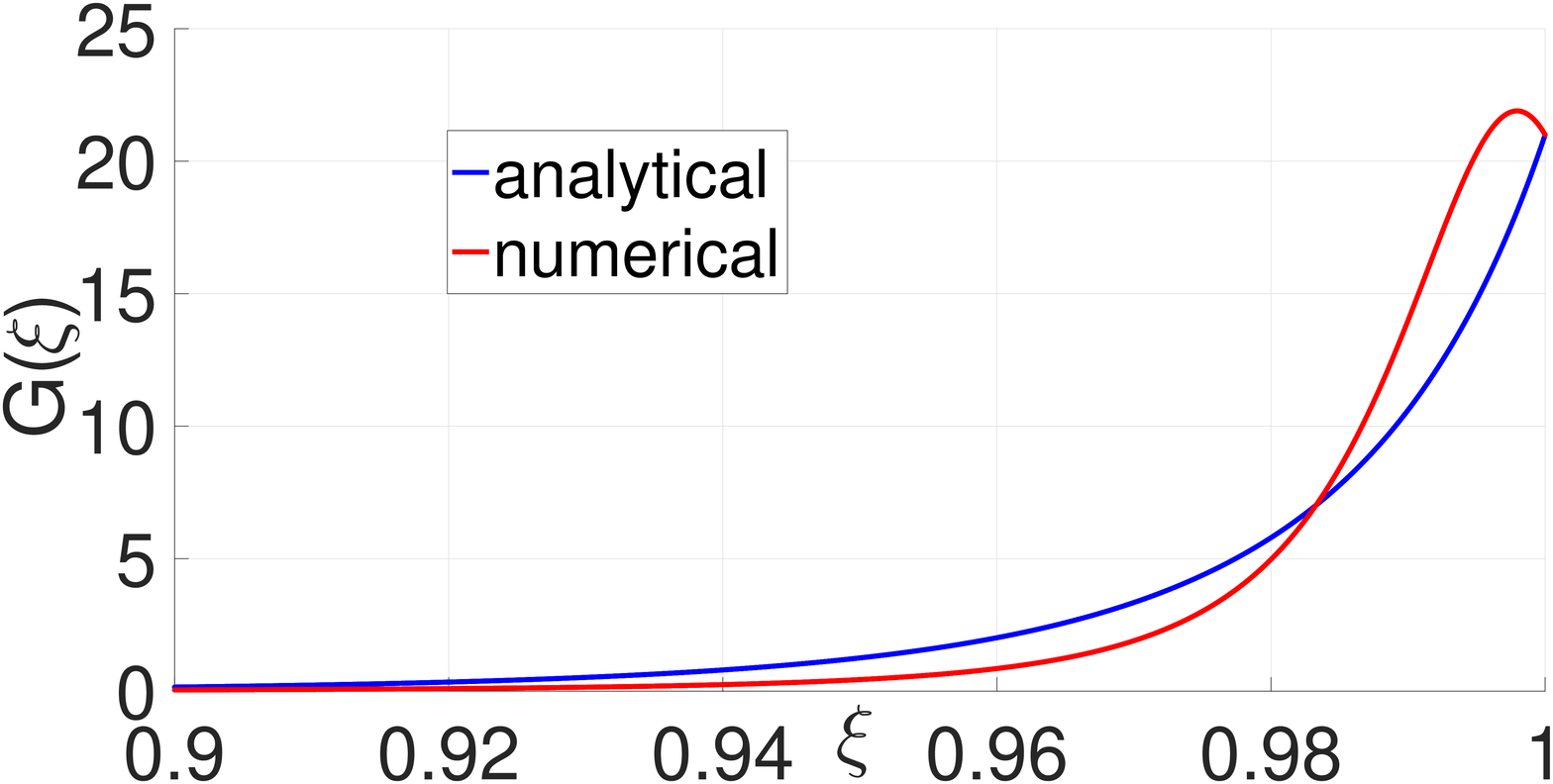} \\ a) $\gamma=1.10$}
\end{minipage}
\hfill
\begin{minipage}[h]{0.5\linewidth}
\center{\includegraphics[width=1.0\linewidth]{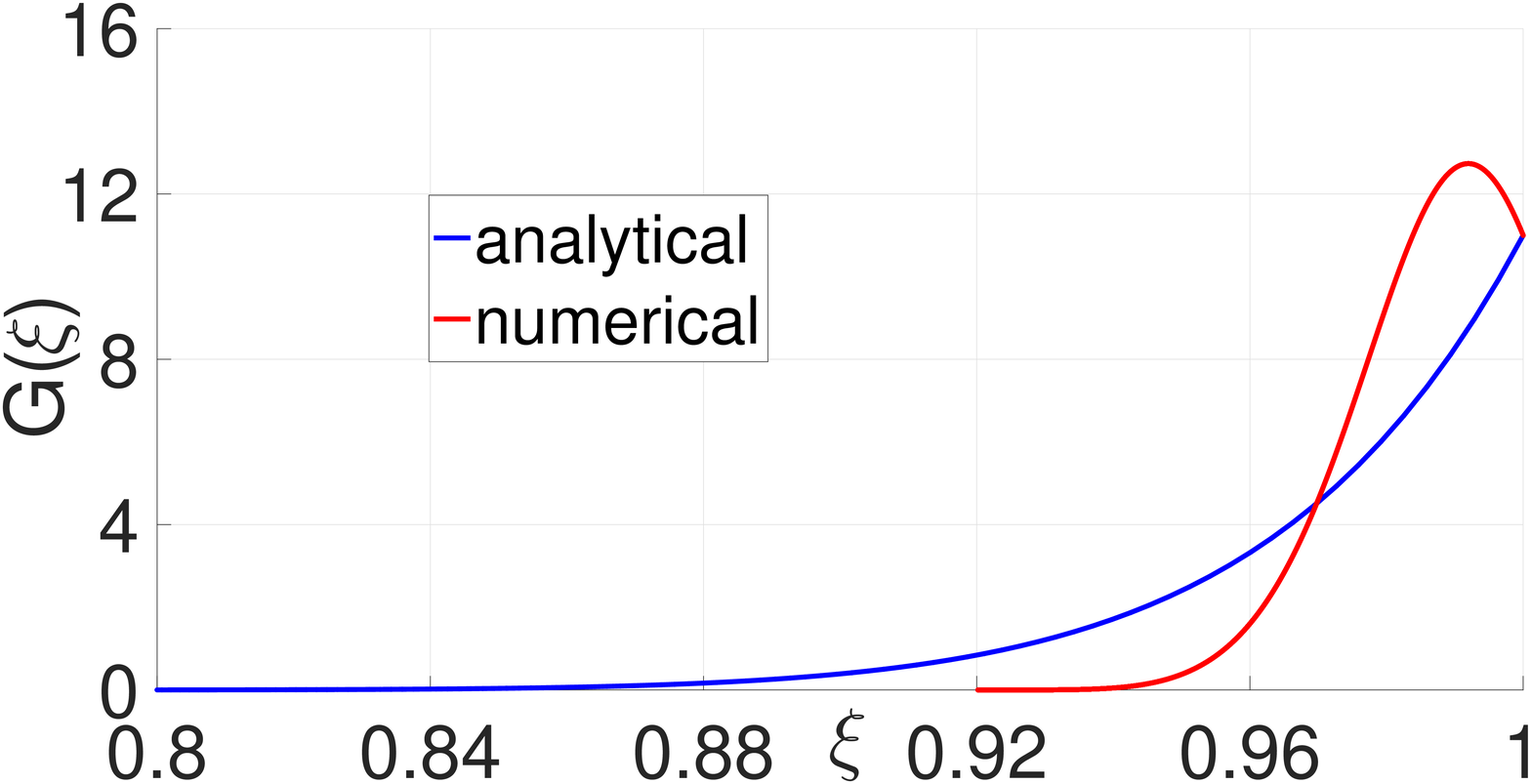} \\b) $\gamma=1.20$}
\end{minipage}
\vfill
\begin{minipage}[h]{0.5\linewidth}
\center{\includegraphics[width=1.0\linewidth]{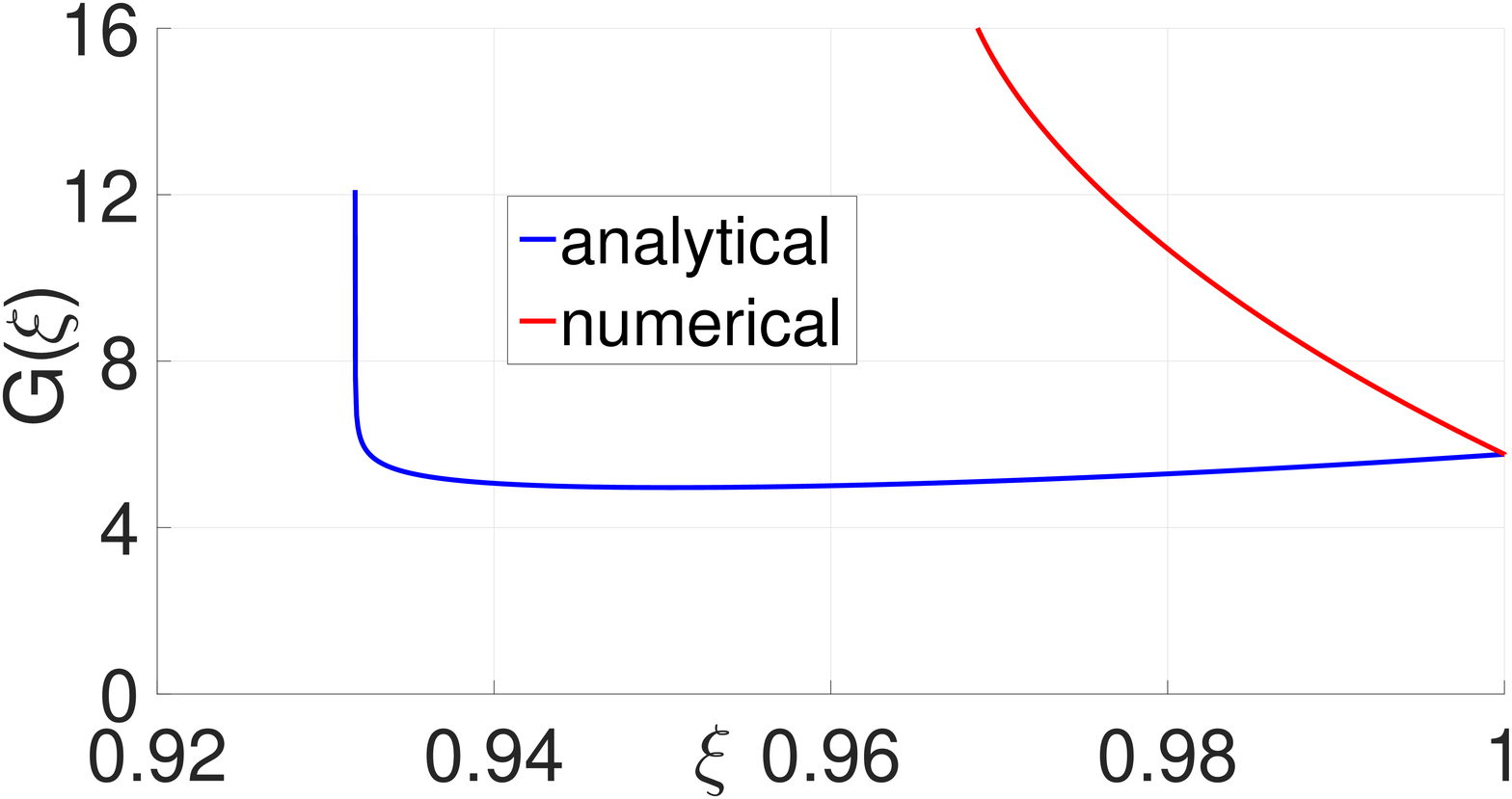} \\c) $\gamma=1.42$}
\end{minipage}
\hfill
\begin{minipage}[h]{0.5\linewidth}
\center{\includegraphics[width=1.0\linewidth]{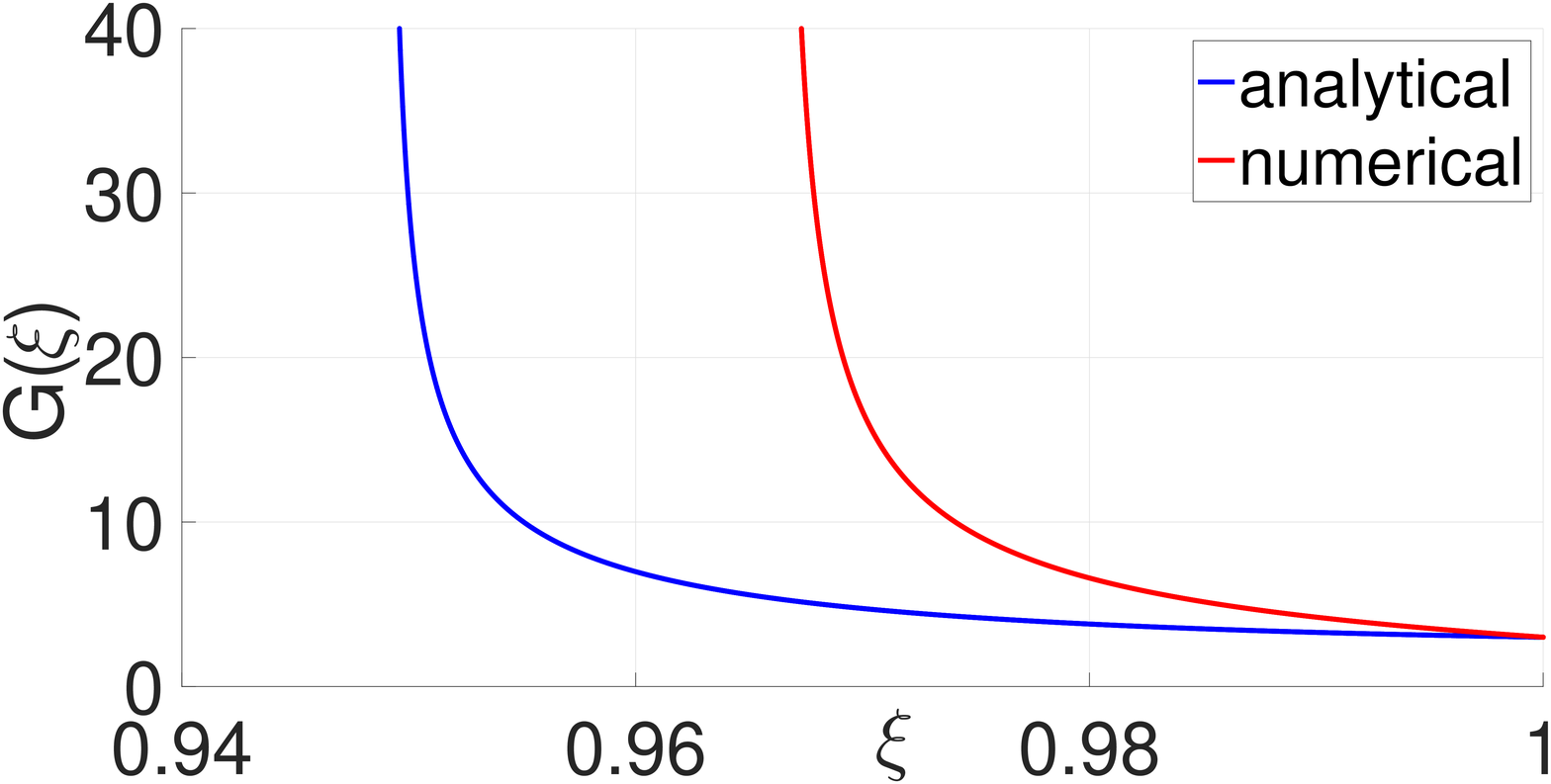} \\ d) $\gamma=2.00$}
\end{minipage}
\caption{Comparison of analytic and numerical curves for $G(\xi)$ at different $\gamma$, in the vicinity of
the shock.
{\bf a.} Example of the case without void, at $1<\gamma<1.1782$ (analytic);  $1<\gamma<1.155$ (numerical).
{\bf b.} Example of the case with void, at $1.1782<\gamma<1.4$ (analytic);  $1.155<\gamma<1.4$ (numerical), when the density at the edge of the void $G(\xi_*)=0$ in both solutions.
{\bf c.} Example of the case with void, at $1.4<\gamma<1.543$ (analytic);  $\gamma>1.4$ (numerical), when the density at the edge of the void $G(\xi_*)=\infty$ in both solutions, and there is a minimum in the analytical curve.
{\bf d.} Example of the case with void, at $\gamma>1.543$ (analytic);  $\gamma>1.543$ (numerical), when the density at the edge of the void $G(\xi_*)=\infty$ in both solutions, and  the analytic curve does not have a minimum.
}
\label{imGannum}
\end{figure}

It may be seen from Fig. \ref{imGannum} that approximate analytical solution for $G(\xi)$ shows all principal layer behavior features. So it is possible to use approximate solution for different estimations.

\begin{figure}
\center{\includegraphics[width=1.0\textwidth]{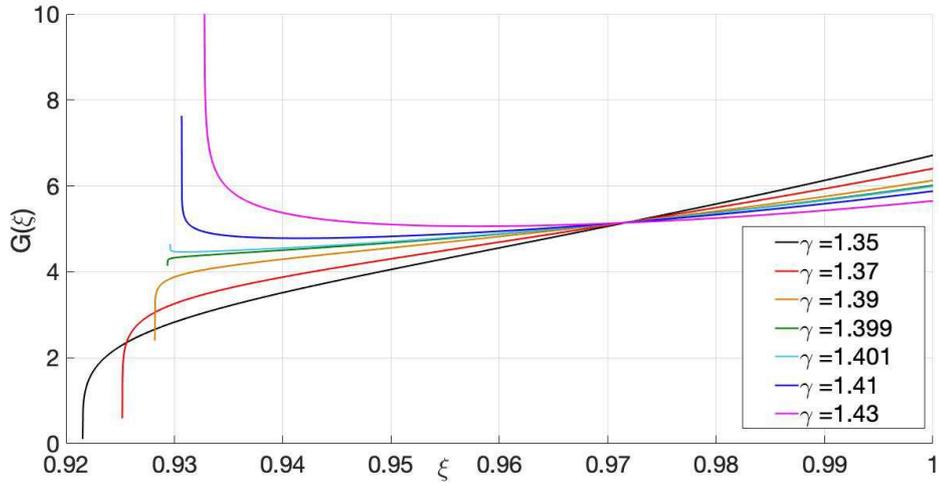}}
\caption{Approximate analytic solution for $G(\xi)$ at $\gamma \approx 1.4$}
\label{im22}
\end{figure}

\begin{figure}
\center{\includegraphics[width=1.0\textwidth]{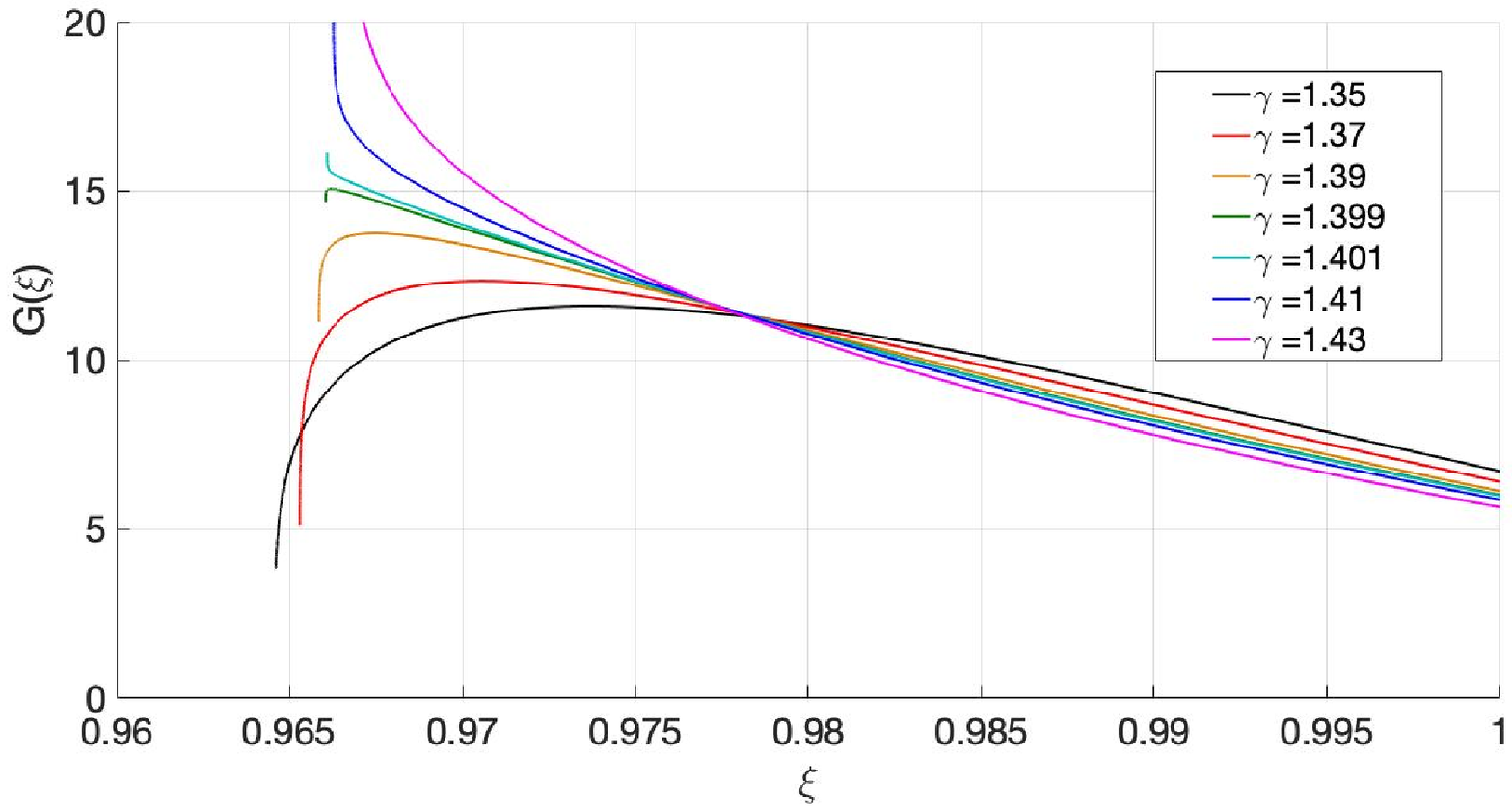}}
\caption{Numerical solution for $G(\xi)$ at $\gamma \approx 1.4$}
\label{im23}
\end{figure}

We have made the high precision calculation and got the results, which are shown in Figs. \ref{im22},\ref{im23}. As we can see the density at the inner edge of the layer is jumping from zero to infinity. Comparing of these figures we have made a conclusion the transition value $\gamma_{cr1}$ is equal to 1.4 at the precision of calculations.

\noindent The constant $\beta$ in the definition of the non-dimensional radius $\xi$ in (\ref{eq24a}) is obtained from the
explosion energy integral $E$. Due to zero energy (kinetic + gravitational) in the non-perturbed solution the conserving value of the explosion energy behind the shock in
the uniformly expanding medium with velocity and density distribu\-ti\-ons (\ref{eq20a}) with account of the gravitational energy determined in (\ref{eq52a})

In non-dimensional variables (\ref{eq24a}) this relation for solutions with hollow center  reduces to the equation for the constant $\beta$
\be
\label{eq53aa}
\beta^{-5}=\frac{64\pi}{25}\int_{\xi_*}^1 G\left[\frac{V^2}{2}+\frac{Z}{\gamma(\gamma-1)}\right]\xi^4 d\xi-\frac{8}{3}\int_{\xi_*}^1 G\xi\left(\int_0^\xi G\eta^2 d\eta\right)d\xi.
\ee

\begin{table}[h]
\caption{The values $\beta(\gamma)$ for the analytic and numerical solutions}
\label{tabular:timesandtenses1}
\begin{center}
\begin{tabular}{ |c |c |c|}
\hline
\textbf{$\gamma$} & \textbf{$\beta_{an}$} & \textbf{$\beta_{num}$} \\
\hline
1.05 &  3.2910 & 3.3512 \\
\hline
1.10 & 2.2268 & 2.5003 \\
\hline
1.12 & 2.0423 & 2.3713 \\
\hline
1.15 & 1.8522 & 2.2416 \\
\hline
1.17 & 1.7631 & 2.1785 \\
\hline
1.20 & 1.6667 & 2.1041 \\
\hline
1.35 & 1.4604 & 1.8897 \\
\hline
1.45 & 1.4048 & 1.8050 \\
\hline
1.60 & 1.3554 & 1.6709 \\
\hline
2.00 & 1.2814 & 1.1298 \\
\hline
\end{tabular}
\end{center}
\end{table}

\noindent The values of $\beta(\gamma)$ for the analytic and numerical  solutions are given in the Table \ref{tabular:timesandtenses1}.
 It follows from numbers in this table, that
 the value of $\xi_*$ has its maximum value both in analytic and numerical models.
 It means that the thickness of the layer goes through the minimum. For $\gamma=10$ the value of $\xi_*^{an}$ is close to its limiting
value in (\ref{42aa}). Actually the results for large $\gamma > \sim 5$, which are obtained from self-similar solution, are not
reliable. At large $\gamma$ the matter compressibility decreases, and the shock is becoming weaker. Hugoniot relations in the
form (\ref{eq23a}) describing the strong shock are not valid anymore. With general Hugoniot adiabatic relations \cite{llhydro}
we cannot construct a self-similar solution. Therefore the results for large $\gamma$ could be considered only as rough
estimations by the order of magnitude.

The high precision calculation for the case of $gamma$ around 1.4, gave the results, which are shown in
Figs. \ref{im22},\ref{im23}. As we can see the density at the inner edge of the layer is jumping from zero to infinity.
Comparing these figures we derive  the transition value of $\gamma_{cr1}$ is equal to 1.4 in both solutions,
within the precision of calculations.

\section*{Acknowledgments}

  This work  was partially supported by
  RFBR grants 18-02-00619, 18-29-21021 and 20-02-00455.

\end{document}